\documentclass[final,3p,times,twocolumn,authoryear]{elsarticle}

\usepackage{amssymb}

\usepackage{amsmath}
\usepackage{amssymb,url,graphicx,booktabs,cite,epstopdf,textcomp,mathrsfs,epsfig,array,setspace}
\usepackage{graphicx}
\usepackage{amsfonts}
\usepackage{amsmath}
\usepackage{placeins}
\usepackage{float}
\restylefloat{table}
\usepackage{verbatim}
\usepackage{multirow}
\usepackage{caption}
\usepackage{subfig}
\usepackage{amsmath}
\usepackage{tikz}
\usepackage{mathdots}
\usepackage{mathtools}
\usepackage{yhmath}
\usepackage{cancel}
\usepackage{color}
\usepackage{siunitx}
\usepackage{array}
\usepackage{multirow}
\usepackage{amssymb}
\usepackage{gensymb}
\usepackage{tabularx}
\usepackage{booktabs}
\usetikzlibrary{fadings}
\usepackage{natbib}
\usepackage[title]{appendix}

\newcommand{\bm}{\begin{bmatrix}}
\newcommand{\fm}{\end{bmatrix}}

\begin{document}

\begin{frontmatter}


\title{Modeling the Role of Land Conversion on the Spread of an Epizootic Disease}
\author[label1]{Dustin G. Padilla}
\ead{dustin.padilla@asu.edu}

\address[label1]{
Simon A. Levin Mathematical, Computational, and Modeling Sciences Center, Arizona State University, 1031 Palm Walk, Tempe, AZ 85287, USA}

\author[label2]{Xiaoqian Gong}

\address[label2]{
School of Mathematical and Statistical Sciences, Arizona State University, 900 Palm Walk, Tempe, AZ 85281, USA }

\author[label3]{Charles Perrings}

\address[label3]{School of Life Sciences and ecoSERVICES Group, Arizona State University, 427 E Tyler Mall,Tempe, AZ 85287, USA}



\begin{abstract}
Land conversion and the resulting contact between domesticated and wild species has arguably been the single largest contributor to the emergence of novel epizootic and zoonotic diseases in the past century. An unintended consequence of these interactions is zoonotic or epizootic disease spillovers from wild species to humans and their domesticates. Disease spillovers are edge effects of land conversion and are sensitive to the size and shape of converted areas. We combine spatial metrics from landscape ecology with theoretical epidemiological models to understand how the size and shape of land conversion affect epizootic and zoonotic disease transmission of single and two species populations. We show that the less compact the converted area, and the the greater the depth of the contact zone, the more rapidly will an introduced disease spread through the domesticated population.
\end{abstract}


\end{frontmatter}




\section{Introduction}

Land conversion has been identified as a primary driver in the emergence and re-emergence of many infectious zoonotic and epizootic diseases \citep{jones2008global, patz2004unhealthy, wolfe2007origins}. When wildlands are converted for anthropogenic purposes, a mosaic of land covers is formed with contact zones between converted and unconverted habitats -- a process known as habitat fragmentation \citep{wilcove1986habitat}. This changes the density, relative abundance, and geographic distribution of both wild species and humans or their domesticates across the landscape \citep{saunders1991biological, wilcove1998quantifying, murcia1995edge}. Contact between wild species and humans or their domesticates at the edges of converted lands then allows for epidemiological spillovers, an ecological edge effect that can lead to the emergence of newly mutated strains of diseases \citep{suzan2012habitat, fahrig2003effects, greer2008habitat}.  The less compact/more fragmented the converted area, the greater are the associated edge effects, including the transmission of novel zoonotic diseases among humans and their domesticates \citep{patz2000effects, daszak2000emerging}.

There is extensive and long-standing empirical research on the relationship between infectious disease, especially vector-borne disease, and habitat fragmentation \citep{giglioli1963ecological,manzione1998venezuelan,estrada2010trend,rulli2017nexus,langlois2001landscape}. Research on how habitat fragmentation and the ensuing edge effects impact zoonotic disease emergence and spread has focused on three topics: (a) changes in fragment biocomposition, (b) the influence of edge effects on species interactions, and (c) disease ecology in isolated areas. It has covered a range of habitat types, matrix compositions, fragment sizes, and landscape histories, as well as three separate fields of inquiry: landscape ecology, mathematical epidemiology, and spatial epidemiology. Nonetheless, the ecological and epidemiological consequences of patterns of conversion and fragmentation on disease dynamics are still poorly understood \citep{suzan2012habitat}. Currently, there is no accepted general theoretical framework for analyzing the impact of habitat fragmentation on zoonotic disease emergence and transmission. Previous attempts to build theory involve either phenomenological models, only relevant in highly specific contexts, or lack essential landscape features \citep{faust2018pathogen,cantrell2001brucellosis,white2018disease}. 

The aim of this paper is to model the relation between landscape structure and disease transmission in order to improve understanding of the connection between habitat fragmentation, and the emergence and spread of zoonotic diseases. Our approach to the issue combines landscape ecology, spatial epidemiology, and mathematical epidemiology. The epidemiological core of the models, described in the next section, is derived from standard mathematical epidemiology \citep{kermack1932contributions, hethcote2000mathematics,brauer2012mathematical}, which uses differential equations to model disease transmission through a population. Models of this sort have already been incorporated into landscape ecology to study disease spread within populations, metapopulations, and communities. They have also been adopted by spatial epidemiologists, who investigate how spatial-temporal environmental variation influences disease risk of populations over a landscape \citep{pavlovsky1966natural}. What we add are the insights obtained from topological landscape ecology, which quantifies landscape heterogeneity through the use of landscape metrics \citep{laurance1991predicting}.


\section{A Model of Wild-to-Domesticated Species Disease Transmission}

To begin, we assume that there are two populations, a population of humans or their domesticates occupying a converted area and a wild population occupying an unconverted area. We abstract the disease and population dynamics of the wild species, and only consider disease and population dynamics in the introduced (domesticated) species. We also assume homogenous mixing within both areas. Land conversion and the introduction of domesticated species generally occur on unconverted land where wild species reside and exhibit their own disease dynamics. Here we assume a wild population to be a disease reservoir, and, as a first approximation, also assume every individual in the wild population to be infectious for all time. Thus, there is unidirectional constant infection from the wild population to domesticated population. The boundary between the two habitats -- the edge -- comprises a contact zone in which the species can interact, and transmit the disease to one another. We later consider disease dynamics within both wild and domesticated populations.

 \citet{hadeler2009epidemic} investigated the introduction of a disease reservoir in classical $SIR$ models based on the premise that even though these models assume an uninfected stationary state, many times this is not the case. Like  \citet{hadeler2009epidemic}, we assume that the wild population is of constant size at equilibrium and is infectious at a constant rate. Unlike \citet{hadeler2009epidemic}, however, we consider how landscape heterogeneity factors into infection. In order to identify the edge effects of land conversion impacts on disease transmission, we calculate the relative size of the contact zone using landscape metrics derived by \citet{patton1975diversity} and \citet{laurance1991predicting} and incorporate this into the estimation of the infection rate from the disease reservoir. This enables us to obtain general results about the impact of the relative size and shape of the converted area and the associated contact zone on disease transmission.

Consider a land area, $a$, lying within a region, $z$, which comprises the portion of $z$ that is converted for the use of human domesticates. We could as well consider a human population, but, to fix ideas, will consider a domesticated species.  The converted area defines the range over which domesticates are distributed. The remainder of the land area, $z-a = b$, is comprised of unconverted wild habitat. Inside area $a$, let there be an area, $o$, that contains all points within a fixed distance, $d$, of the edge of $a$ and $b$. We suppose that the range of wild species is $b + o$. The domesticated population, of size $N_D$, is assumed to be evenly distributed within $a$, and the wild population, of size $N_W$, is assumed to be evenly distributed within $b + o$. The area $o$ is the contact zone where an overlap or mixing of the populations $N_D$ and $N_W$ occurs. The core habitat of $N_D$, is denoted $c = a - o$. The ratio of $o$ to $c$ depends on the distance that wild species are able to penetrate the converted area which gives the depth of the contact zone, $d$, and the compactness of $a$. The depth of the contact zone might depend on species characteristics. For example, mosquitoes and bats differ in the distance they are able to fly to feed on the domesticated population. It might also depend on measures taken by people to secure the edge of converted areas, such as the construction of veterinary cordon fences or other barriers.

Let the domesticated population, $N_D$, be compartmentalized into classes $S_D$, for susceptible individuals, $I_D$, for infectious individuals, and $R_D$, for recovered individuals, such that $N_D=S_D+I_D+R_D$. Let the birth rate of susceptible individuals be $\Lambda$, and let deaths in all compartments occur at rate $\sigma$ for population $N_D$ so that $N_D'=\Lambda - \sigma N_D$. Moreover, $N_D > 0$, and $S_D\ge0$, $I_D\ge0$, and $R_D\ge0$ for all time $t$.

\begin{figure}[!htp]
    \centering
\tikzset{every picture/.style={line width=0.75pt}} 

\begin{tikzpicture}[x=0.4pt,y=0.4pt, yscale=-1, xscale=1]

\draw   (127,213.5) .. controls (127,128.72) and (195.72,60) .. (280.5,60) .. controls (365.28,60) and (434,128.72) .. (434,213.5) .. controls (434,298.28) and (365.28,367) .. (280.5,367) .. controls (195.72,367) and (127,298.28) .. (127,213.5) -- cycle ;
\draw   (173.13,213.5) .. controls (173.13,154.2) and (221.2,106.13) .. (280.5,106.13) .. controls (339.8,106.13) and (387.88,154.2) .. (387.88,213.5) .. controls (387.88,272.8) and (339.8,320.88) .. (280.5,320.88) .. controls (221.2,320.88) and (173.13,272.8) .. (173.13,213.5) -- cycle ;
\draw    (127,213.5) -- (173.13,213.5) ;
\draw   (23,98) .. controls (23,55.47) and (57.47,21) .. (100,21) -- (461,21) .. controls (503.53,21) and (538,55.47) .. (538,98) -- (538,329) .. controls (538,371.53) and (503.53,406) .. (461,406) -- (100,406) .. controls (57.47,406) and (23,371.53) .. (23,329) -- cycle ;

\draw (144,220) node [anchor=north west][inner sep=0.75pt]  [font=\normalsize]  {$d$};
\draw (269,115) node [anchor=north west][inner sep=0.75pt]  [font=\normalsize]  {$c$};
\draw (269,65) node [anchor=north west][inner sep=0.75pt]  [font=\normalsize]  {$o$};
\draw (269,25) node [anchor=north west][inner sep=0.75pt]  [font=\normalsize]  {$b$};
\end{tikzpicture}
\caption{An illustration of the habitat layout for the domesticated and wild species. The domesticated population is assumed to inhabit the core habitat, $c$, and the contact zone, $o$. The wild population is assumed to inhabit the unconverted wild habitat, $b$, and the contact zone, $o$. Note that the converted habitat is $a=c+o$. $d$ is the contact zone depth that the wild population penetrates into the domesticated population's habitat.}
\label{fig:layout}
\end{figure}
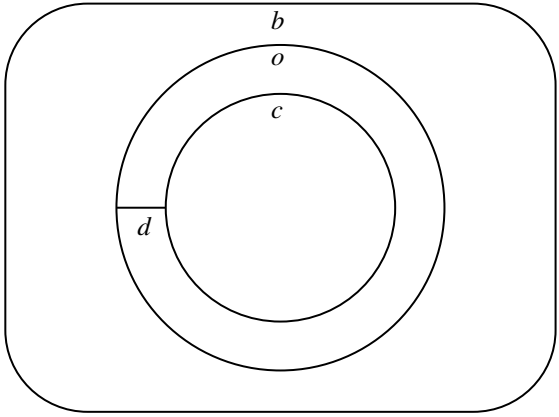



Following the technique of \citet{brauer2017final}, we define the transmission rates in a way such that the system of equations reduces to mass action. Let $b_{WD}$ be the average number of cross-species contacts that a wild animal makes in unit time, and let $f_{WD}$ be the probability that cross-species contact between a wild infectious individual and a susceptible domesticated individual transmits infection. The transmission rate of the disease from wild to domesticated animals is the probability that a cross-species contact results in infection times the number of contacts in unit time, \[\delta_D=b_{WD}f_{WD}N_W.\] 
 A susceptible domesticate then has $\delta_D$ infective interactions with the wild population in unit time, of which a fraction $I_W$ is with an infectious wild individual. Thus, the number of new infective domesticated animals caused by infection from a wild animal in unit time is
\[\delta_D S_D\frac{I_W}{N_W}.\]
As a first approximation, there is a constant infection risk from wild species in the contact zone. We suppose that the entire wild population is infectious, $I_W=N_W$. The size of the domesticated population that is exposed at any moment depends on the size of the contact zone relative to the converted area. 
Let $b_{DD}$ be the average number of within-species contacts that a domesticated animal makes in unit time, and let $f_{DD}$ be the probability that cross-species contact between a domesticated infectious individual and a susceptible domesticated individual transmits infection. Hence, the transmission rate of the disease from one domesticated animal to another is the probability that that within-species contact results in an infection times the number of contacts had by an average domesticate in unit time. The number of newly infectious domesticated animals caused by infection from a domesticated animal in unit time is 
\[\beta_D=b_{DD}f_{DD}N_D.\]
Infected individuals are assumed to recover at rate $\gamma$, and lose immunity at rate $\eta$. For the numerical simulations, we assume the initial number of infectious individuals to be at equilibrium, calculated from the two-species epidemic model model in Section 3. This describes how a disease-free population responds when it is introduced to another population with an endemic disease. Since the populations are assumed to be evenly distributed throughout their respective habitats, $\frac{o}{a}$ is the proportion of $S_D$ that is in $o$ at any given time, and $\frac{o}{b+o}$  is the proportion of the wild population that is in $o$ at any given time. Recognizing that $b=z-a=z-c-o$, the model can be written as

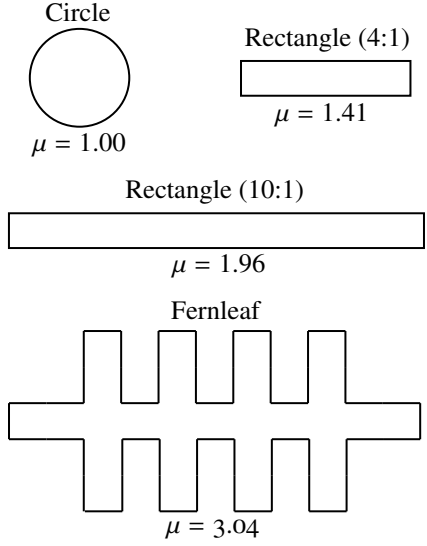
\begin{figure}[!htp]\centering
\begin{center}

\tikzset{every picture/.style={line width=0.75pt}}       

\begin{tikzpicture}[x=0.75pt,y=0.75pt,yscale=-1,xscale=1]

\draw   (198.13,53.25) .. controls (198.13,39.66) and (209.24,28.65) .. (222.95,28.65) .. controls (236.66,28.65) and (247.78,39.66) .. (247.78,53.25) .. controls (247.78,66.84) and (236.66,77.86) .. (222.95,77.86) .. controls (209.24,77.86) and (198.13,66.84) .. (198.13,53.25) -- cycle ;

\draw   (303.55,44.79) -- (388.13,44.79) -- (388.13,62.28) -- (303.55,62.28) -- cycle ;

\draw    (225.06,198.77) -- (225.06,216.91) ;

\draw    (243.72,198.77) -- (243.72,216.91) ;

\draw    (225.06,180.63) -- (243.72,180.63) ;

\draw    (225.06,235.05) -- (225.06,253.18) ;

\draw    (187.74,216.91) -- (187.74,235.05) ;

\draw    (281.62,235.05) -- (300.28,235.05) ;

\draw    (206.4,235.05) -- (225.06,235.05) ;

\draw    (206.4,216.91) -- (225.06,216.91) ;

\draw    (243.72,216.91) -- (262.38,216.91) ;

\draw    (244.3,235.05) -- (262.96,235.05) ;

\draw    (262.96,271.32) -- (281.62,271.32) ;

\draw    (281.04,216.91) -- (299.7,216.91) ;

\draw    (225.64,271.32) -- (244.3,271.32) ;

\draw    (187.74,216.91) -- (206.4,216.91) ;

\draw    (187.74,235.05) -- (206.4,235.05) ;

\draw    (262.38,180.63) -- (281.04,180.63) ;

\draw    (225.06,180.63) -- (225.06,198.77) ;

\draw    (243.72,180.63) -- (243.72,198.77) ;

\draw    (225.06,253.18) -- (225.06,271.32) ;

\draw    (244.3,235.05) -- (244.3,253.18) ;

\draw    (244.3,253.18) -- (244.3,271.32) ;

\draw    (262.96,235.05) -- (262.96,253.18) ;

\draw    (262.96,253.18) -- (262.96,271.32) ;

\draw    (262.38,198.77) -- (262.38,216.91) ;

\draw    (262.38,180.63) -- (262.38,198.77) ;

\draw    (281.04,180.63) -- (281.04,198.77) ;

\draw    (281.04,198.77) -- (281.04,216.91) ;

\draw    (281.62,235.05) -- (281.62,253.18) ;

\draw    (281.62,253.18) -- (281.62,271.32) ;

\draw    (299.7,198.77) -- (299.7,216.91) ;

\draw    (299.7,180.63) -- (299.7,198.77) ;

\draw    (300.28,235.05) -- (300.28,253.18) ;

\draw    (300.28,253.18) -- (300.28,271.32) ;

\draw    (318.94,235.05) -- (318.94,253.18) ;

\draw    (318.94,253.18) -- (318.94,271.32) ;

\draw    (318.35,198.77) -- (318.35,216.91) ;

\draw    (318.35,180.63) -- (318.35,198.77) ;

\draw    (355.67,180.63) -- (355.67,198.77) ;

\draw    (355.67,198.77) -- (355.67,216.91) ;

\draw    (356.25,235.05) -- (356.25,253.18) ;

\draw    (356.25,253.18) -- (356.25,271.32) ;

\draw    (337.01,180.63) -- (337.01,198.77) ;

\draw    (337.01,198.77) -- (337.01,216.91) ;

\draw    (337.6,235.05) -- (337.6,253.18) ;

\draw    (337.6,253.18) -- (337.6,271.32) ;

\draw    (300.28,271.32) -- (318.94,271.32) ;

\draw    (299.7,180.63) -- (318.35,180.63) ;

\draw    (318.35,216.91) -- (337.01,216.91) ;

\draw    (337.01,180.63) -- (355.67,180.63) ;

\draw    (337.6,271.32) -- (356.25,271.32) ;

\draw    (318.94,235.05) -- (337.6,235.05) ;

\draw    (356.25,235.05) -- (374.91,235.05) ;

\draw    (355.67,216.91) -- (374.33,216.91) ;

\draw    (374.33,216.91) -- (392.99,216.91) ;

\draw    (374.91,235.05) -- (392.99,235.05) ;

\draw    (392.99,216.91) -- (392.99,235.05) ;

\draw   (187.71,121.4) -- (394.88,121.4) -- (394.88,138.89) -- (187.71,138.89) -- cycle ;

\draw (289.59,280.37) node [scale=1] [align=left] {$\displaystyle {\displaystyle \mathnormal{\mu =3.04}}$};

\draw (290.9,170) node [scale=1] [align=left] {Fernleaf};

\draw (292.57,148) node  [align=left] {$\displaystyle \mu =1.96$};

\draw (290.43,110.8) node [scale=1] [align=left] {Rectangle (10:1)};

\draw (222.22,20) node [scale=1] [align=left] {Circle};

\draw (222.83,86.94) node  [align=left] {$\displaystyle \mu =1.00$};

\draw (346.67,33.92) node [scale=1] [align=left] {Rectangle (4:1)};

\draw (343.96,72) node  [align=left] {$\displaystyle \mu =1.41$};

\end{tikzpicture}

\end{center}

\caption{Examples of shapes with calculated shape indices \citep{laurance1991predicting}.}

\label{fig:shapes}

\end{figure}

\footnotesize{
\begin{equation}\label{eq1}
\left.\begin{aligned}
&S_D' = \Lambda - \delta S_D \left(\dfrac{o}{a}\right)\left(\dfrac{I_W}{N_W}\right)\left(\dfrac{o}{z-c}\right) -\beta S_D\left(\dfrac{I_D}{N_D}\right)+ \eta R_D -\sigma S_D,\\
&I_D' =  \delta S_D \left(\dfrac{o}{a}\right)\left(\dfrac{I_W}{N_W}\right)\left(\dfrac{o}{z-c}\right) +\beta S_D\left(\dfrac{I_D}{N_D}\right)- (\gamma  + \sigma) I_D,\\
&R_D' =  \gamma I_D - (\eta + \sigma) R_D,\\
&S_D(0) = \text{ }N_D(0),\text{ } I_D(0)\text{ } =\text{ }0, \text{ }R_D(0) \text{ }=\text{ } 0, \text{ }N_D \text{ }=\text{ } S_D+I_D+R_D
\end{aligned}\right.
\end{equation}
}
\normalsize
To understand edge effects on disease transmission, we need to be more precise about the factors determining the relative size of the contact zone, $o$. Three factors are critical: the size of the converted area, the compactness of the converted area, and the depth of the contact zone. From \citet{laurance1991predicting}, the size of $o$ can be approximated by the empirically derived equation \[o = kd\mu\sqrt{a}     \text{ }\text{ }\text{, where}\text{ }\text{ } \mu = \dfrac{\ell}{2\sqrt{\pi a}},\] $\ell$ is the perimeter length of the fragmented area $a$, $k=3.55$ is an empirically calculated constant, and $\pi=3.1415$ is the ratio of a circle's circumference to its diameter. $\mu$ is referred to as the ``shape index" in the landscape ecology literature. $\mu$ takes a value of 1 if the shape is a circle (associated with the smallest edge to area ratio) i.e. the most compact shape, and increases in value as the shape is progressively deformed through fragmentation \citep{patton1975diversity}. Figure \ref{fig:shapes} illustrates a variety of shapes and their corresponding shape indices.

The constant $k$ was determined by setting the total area of the domesticated species habitat, $a$, of multiple shapes to 10,000 hectares (ha) with $d = 1$m, and then estimating the rate at which the size of the core area, $c$, of each shape declined as $d$ increased \citet{laurance1991predicting}. A circle of 10,000ha was found to lose 3.55ha for every 1m increase in $d$. The decline in the core area/increase in the edge was found to be nearly linear.  As the depth of the contact zone increased, the size of the core decreased at a constant shape-specific rate, up to a minimum value that was approximately $50\%$ of $a$ \citep{laurance1991predicting}. The rate at which an increase in the depth of the contact zone affected the size of the core was, however, sensitive to the compactness of the converted area.  Less compact/ more irregularly shaped areas were shown to accrue edge effects more rapidly than more compact/ less irregular areas.

Substituting the approximated size of the overlap into the model and letting \[\delta_1=\delta\left[\frac{(kd\mu)^2}{z-c}\right]\] the system of equations becomes

\footnotesize
\begin{equation}\label{eq2}
\left.\begin{aligned}
&S_D' = \Lambda- \delta_1S_D\left(\dfrac{I_W}{N_W}\right) -\beta S_D\left(\dfrac{I_D}{N_D}\right)+ \eta R_D -\sigma S_D,\\
&I_D' =  \delta_1S_D\left(\dfrac{I_W}{M}\right) +\beta S_D\left(\dfrac{I_D}{N_D}\right)- (\gamma + \sigma) I_D,\\
&R_D' =  \gamma I_D- (\eta + \sigma) R_D,\\
&S_D(0) = \text{ }N_D(0),\text{ } I_D(0)\text{ } =\text{ }0, \text{ }R_D(0) \text{ }=\text{ } 0, \text{ }N_D \text{ }=\text{ } S_D+I_D+R_D
\end{aligned}\right.
\end{equation}
\normalsize



\subsection{Epidemiological Characteristics}



If the equations in (\ref{eq2}) are added, the differential equation describing the change of the total population size is
\[
N_D' = \Lambda - \sigma N_D,
\]
for which the solution is 
\[N_D(t)=(\Lambda/\sigma)+N_D(0)e^{( -\sigma t)}.\]
As $t\rightarrow\infty$, then $N_D\rightarrow\Lambda/\sigma$. Since the population is asymptotically constant, $S_D=(\Lambda/\sigma)-I_D-R_D$, by \citet{castillo1994asymptotically}, system (2) is reduced to
%
%
\begin{equation}\label{eq3}
\left.\begin{aligned}
&I_D' =  \delta_2 (K- I_D -R_D) + \beta (K- I_D -R_D) I_D/K - (\gamma + \sigma)I_D,\\\
&R_D' =  \gamma I_D - (\eta+ \sigma) R_D,
\end{aligned}\right.
\end{equation}
where $\delta_2=\delta_1(I_W/N_W)$ and $K_D=\Lambda/\sigma$.
The system has one biologically relevant (positive) equilibrium $(I_D^{*},R_D^{*})$, and is globally stable by Dulac's criterion and the Poincar\'e-Bendixson theorem.

Since there are two sources of infection, $\delta_1$ (corresponding to wild infections which are assumed constant within the contact zone), and $\beta$ (corresponding to domestic infections), it is necessary to consider cases to understand the contagion process.

 If $\delta_1=0$, then we have the classical $SIR$ model, and the condition for $I(t)$ to be increasing is $\beta>\gamma$; thus, $\mathcal{R}_0=\beta/\gamma$.  If $\delta_1>0$ and $\beta=0$, then the wild species can infect the domesticated species, but there is no transmission within the domesticated species, such as with mosquito-borne diseases. Moreover, for any $\delta_1>0$, $I(t)$ will always be increasing, and is a sufficient condition for the spread of the disease. All solutions converge to the same interior equilibrium and the characterization of $\mathcal{R}_0$ is not valid. Disease prevalence, the number of infected individuals at a particular time, can be reduced by changing the parameter values associated with land conversion $\mu$, $d$, and $\ell$ of system (\ref{eq3}).

\section{A Model of Disease Transmission between Domesticated and Wild Populations}

We now relax the assumption that disease transmission occurs only in one direction -- from wild to domesticated populations -- in order to understand how landscape features affect overall prevalence. Again, assume there are two populations situated on their respective habitats, with a contact zone in-between where the two populations can interact. Consider the domesticated population to be compartmentalized into disease state classes as previously, and, similarly, let the wild population, $N_W$, be separated into compartments $S_W$, for susceptible wild individuals, $I_W$, for infectious wild individuals, and $R_W$ for recovered wild individuals .

%


For the domesticated population $N_D$, new births are susceptible, denoted $\Lambda_D$, and natural mortality occurs in all compartments at rate $\sigma_D$, so that $N_D'=\Lambda_D - \sigma_D N_D$. Likewise, for the wild population $N_W$, all new births are assumed to be susceptible, denoted $\Lambda_W$, and deaths occur in all compartments at rate $\sigma_W$, so that $N_W'=\Lambda_W - \sigma_W N_W$. Moreover, $N_D > 0$, and $S_D\ge0$, $I_D\ge0$, and $R_D\ge0$ for all time $t$. As well as and $N_W > 0$, and $S_W\ge0$, $I_W\ge0$, and $R_W\ge0$ for all time $t$. 

Assume that transmission of the disease to the domesticated population occurs when a susceptible domesticated animal interacts with an infectious wild animal and there is successful transmission.

 Let $b_{WD}$ be the average number of cross-species contacts that a wild animal makes in unit time, and let $f_{WD}$ be the probability that cross-species contact between a wild infectious individual and susceptible domesticated individual transmits infection. The transmission rate of the disease from wild to domesticated animals is then the probability a cross-species contact transmits infection times the number of contacts in unit time,
\[
\delta_D=b_{WD}f_{WD}N_W.
\]
A susceptible domesticated animal has $\delta_D$ infective interactions with the wild population in unit time, of which a fraction $I_W$ is with infectious wild animals. The number of new infective domesticated animals caused by infection from a wild animal in unit time is
\[
\delta_DS_D\dfrac{I_W}{N_W}.
\]
Similarly, the transmission rate of the disease from domesticated to wild animals is the probability that that cross-species contact transmits infection times the number of contacts in unit time,
\[
\delta_W=b_{DW}f_{DW}N_D.
\]
A susceptible wild animal has $\delta_D$ infective interactions with the domesticated population in unit time, of which a fraction $I_D$ is with an infectious domesticated individual. The number of newly infected wild animals caused by infection from a domesticated animal in unit time is then
\[
\delta_{DW}S_W\dfrac{I_D}{N_D}.
\]
Within-species infections are modeled in a similar way. Specifically, let $b_{DD}$ be the average number of contacts that domesticated animals make with their own kind in unit time, and let $f_{DD}$ be defined as the probability that contact of domesticated animals with their own kind transmits infection. The transmission rate is the probability that a contact transmits infection multiplied by the number of within-species contacts in unit time
\[
\beta_D=b_{DD}f_{DD}N_D.
\]
The within-species transmission rate for wild animals is defined in a similar way such that 
\[
\beta_W=b_{WW}f_{WW}N_W.
\]
After a successful disease transmission, the formerly pathogen-free individual becomes infectious and is able to infect other susceptible domesticated individuals at rate $\beta_D$. Infectious domesticated individuals die from the illness at rate $\sigma_D$, recover at rate $\gamma_D$ and lose temporary immunity at rate $\eta_D$. We assume, also, that the disease can be transmitted to the susceptible wild population either through contact with infectious wild individuals at rate $\beta_W$, or through contact with infectious domesticated individuals within the contact zone, becoming infectious at rate $\delta_W$. Infectious wild individuals are assumed to recover at rate $\gamma_W$ and lose temporary immunity at rate $\eta_W$. The size of the contact zone, $o$, can again be approximated using the empirically derived formulations \citep{laurance1991predicting}. Based on these assumptions, a bidirectional zoonotic disease transmission model that incorporates edge effects is

\scriptsize
\begin{equation}\label{eq4}
\left.\begin{aligned}
&S_D' = \Lambda_D  - \delta_D\left[\dfrac{(kd\ell)^2}{4\pi a (z-c)}\right]S_D\dfrac{I_W}{N_W} -\beta_D S_D\left(\dfrac{I_D}{N_D}\right)-\sigma_DS_D +\eta_DR_D,\\
&I_D' = \delta_D\left[\dfrac{(kd\ell)^2}{4\pi a (z-c)}\right]S_D\dfrac{I_W}{N_W}+\beta_D S_D\left(\dfrac{I_D}{N_D}\right)-(\sigma_D +\gamma_D)I_D,\\
&R_D' = \gamma_D I_D- (\sigma_D + \eta_D)R_D,\\
&S_W' = \Lambda_W -\delta_W\left[\dfrac{(kd\ell)^2}{4\pi a (z-c)}\right]S_W\dfrac{I_D}{N_D} -\beta_W S_W\left(\dfrac{I_W}{N_W}\right)-\sigma_WS_W +\eta_WR_W,\\
&I_W' = \delta_W\left[\dfrac{(kd\ell)^2}{4\pi a (z-c)}\right]S_W\dfrac{I_D}{N_D} + \beta_W S_W\left(\dfrac{I_W}{N_W}\right)-(\sigma_W+\gamma_W)I_W,\\
&R_W' = \gamma_W I_W - (\sigma_W + \eta_W) R_W,\\
&S_D(0) = \text{ }N_D(0),\text{ } I_D(0)\text{ } =\text{ }0, \text{ }R_D(0) \text{ }=\text{ } 0, \text{ }N_D \text{ }=\text{ } S_D+I_D+R_D,\\
&S_W(0) = \text{ }N_W(0),\text{ } I_W(0)\text{ } =\text{ }0, \text{ }R_W(0) \text{ }=\text{ } 0, \text{ }N_W \text{ }=\text{ } S_W+I_W+R_W
\end{aligned}\right.
\end{equation}
\normalsize



\subsubsection{Model Characteristics}




If the equations in (\ref{eq4}) are added, the differential equations describing the change of the total population sizes are obtained
\[
N_D' = \Lambda_D - \sigma_D N_D,
\]
\[
N_W'= \Lambda_W -\sigma_W N_W,
\]
for which the solutions are
\[N_D(t)=(\Lambda_D/\sigma_D)+N_D(0)e^{( -\sigma_D t)},\]
\[N_W(t)=(\Lambda_W/\sigma_W)+N_W(0)e^{( -\sigma_W t)}.\]
As $t\rightarrow\infty$, then $N_D\rightarrow\Lambda_D/\sigma_D$ and $N_W\rightarrow\Lambda_W/\sigma_W$. Letting $K_D=\Lambda_D/\sigma_D$ and $K_W=\Lambda_W/\sigma_W$, it can be seen that system (\ref{eq4}) has the disease free equilibrium
\[
E_1=(K_D,0,0,K_W,0,0),
\]
which always exists and is locally asymptotically stable if

\footnotesize
\[
\frac{\beta_D+\beta_W+\sqrt{(\beta_D+\gamma_W+\sigma_W-\beta_W-\gamma_D-\sigma_D)^2+4{\Psi_D}{\Psi_W}}}{\gamma_D+\sigma_D+\gamma_W+\sigma_W}<1
\]
\normalsize
is satisfied, for \[\Psi_i=\delta_i\left[\dfrac{(kd\ell)^2}{4\pi a (z-c)}\right],\] and $i=D,W$. The condition for stability can be rewritten as
\small
\begin{align*}
&\beta_D-\gamma_D-\sigma_D+\beta_W-\gamma_W-\sigma_W\\
&\quad+\sqrt{(\beta_D+\gamma_W+\sigma_W-\beta_W-\gamma_D-\sigma_D)^2+4{\Psi_D}{\Psi_W}}<0.
\end{align*}
\normalsize
We know that 
\small
\begin{align*}
&\beta_D-\gamma_D-\sigma_D+\beta_W-\gamma_W-\sigma_W\\
&\quad+\sqrt{(\beta_D+\gamma_W+\sigma_W-\beta_W-\gamma_D-\sigma_D)^2+4{\Psi_D}{\Psi_W}}<\\
&\beta_D-\gamma_D-\sigma_D+\beta_W-\gamma_W-\sigma_W\\
&+\sqrt{(\beta_D+\gamma_W+\sigma_W-\beta_W-\gamma_D-\sigma_D)^2}+\sqrt{4{\Psi_D}{\Psi_W}}=
\end{align*}
\[
2(\beta_D-\gamma_D-\sigma_D)+\sqrt{4{\Psi_D}{\Psi_W}},
\]
\normalsize
which means that
\[
|\gamma_D+\sigma_D| \ge |2(\beta_D + \sqrt{{\Psi_D}{\Psi_W}})|
\]
is sufficient for the stability condition to be satisfied. However, if 
\[
|\gamma_D+\sigma_D| < |2(\beta_D + \sqrt{{\Psi_D}{\Psi_W}})|,
\]
then
\[
|\gamma_D+\sigma_D+\gamma_W+\sigma_W| \ge |2(\beta_D + \beta_W+\sqrt{{\Psi_D}{\Psi_W}})|
\]
is required for the stability condition to be satisfied. This means that for the populations to reach a disease free state, either the removal rates from infectivity of one population must be greater than the rates of cross-species infection and same-species infection, or the removal rates from infectivity of both population must be greater than the rates of cross-species infection and same-species infection. 

If the land conversion parameters are such that neither condition is satisfied then the disease-free equilibrium will be unstable. If one of the land conversion parameters is zero, then the sum of the removal rates have to be greater than the within-species infection rate for the disease-free equilibrium to be stable. Moreover, if the sum of the removal rates equals the within-species infection rates, $\beta_D=\gamma_D+\sigma_D$ and $\beta_W=\gamma_W+\sigma_W$, then the condition for stability is not satisfied, and the disease-free equilibrium is unstable.


The basic reproductive number, the number of new infections caused by a single infectious individual in a completely susceptible population, is calculated directly using the next generation matrix method\citep{van2002reproduction,diekmann1990definition}, and is
%

\[
\mathcal{R}_0=\frac{\Gamma+\sqrt{(-\Gamma)^2+4\Psi_W\Psi_D\Phi}}{2\Phi},
\]

where $\Gamma=\beta_D(\gamma_W + \sigma_W) + \beta_W(\gamma_D + \sigma_D)$, and $\Phi=(\gamma_W + \sigma_W)(\gamma_D + \sigma_D)$.
If $\mathcal{R}_0<1$, then the disease-free equilibrium is stable, and if $\mathcal{R}_0>1$, then the disease-free equilibrium is unstable. We can see that if one of the land conversion parameters, either $\mu$ or $d$, is zero, then the basic reproductive number reduces to an $\mathcal{R}_0$ for a single population,

\[
\mathcal{R}_0=\frac{\beta_D}{\gamma_D+\sigma_D}.
\]

Moreover, as the land conversion parameters increase, the basic reproductive number will increase to a value greater than one, the disease-free equilibrium will be unstable, meaning that disease prevalence will increase.

To determine the existence and stability of any endemic equilibria, and since the populations are asymptotically constant, let

\[S_D=(\Lambda_D/\sigma_D)-I_D-R_D,\] 
and  
\[S_W=(\Lambda_W/\sigma_W)-I_W-R_W,\]
 so that system (\ref{eq4}) is reduced to
\scriptsize
\begin{equation}\label{eq5}
\left.\begin{aligned}
&I_D' = \Psi_D(K_D-I_D-R_D)\dfrac{I_W}{K_W}+\beta_D (K_D-I_D-R_D)\left(\dfrac{I_D}{K_D}\right)-(\sigma_D +\gamma_D)I_D,\\
&R_D' = \gamma_D I_D- (\sigma_D + \eta_D)R_D,\\\
&I_W' = \Psi_W(K_W-I_W-R_W)\dfrac{I_D}{K_D} + \beta_W (K_W-I_W-R_W)\left(\dfrac{I_W}{K_W}\right)-(\sigma_W+\gamma_W)I_W,\\
&R_W' = \gamma_W I_W - (\sigma_W + \eta_W) R_W,\\
&I_D(0)\text{ } =\text{ }0, \text{ }R_D(0) \text{ }=\text{ } 0, \text{ }N_D \text{ }=\text{ } S_D+I_D+R_D,\\
&I_W(0)\text{ } =\text{ }0, \text{ }R_W(0) \text{ }=\text{ } 0, \text{ }N_W \text{ }=\text{ } S_W+I_W+R_W.
\end{aligned}\right.
\end{equation}
\normalsize
Solving for the equilibria of (\ref{eq5}) yields the following cubic polynomial
\begin{equation}\label{eq6}
\tilde{A}I_W^3+\tilde{B}I_W^2+\tilde{C}I_W+\tilde{D}=0,
\end{equation}
in which
\begin{equation}
\left.\begin{aligned}
&\tilde{A} = a_1 (a_2)^2 - a_2 b_1 b_2, \nonumber\\
&\tilde{B} = a_2 b_2 c_1 - 2 a_1 a_2 c_2 + b_1 b_2 c_2 + (b_2)^2 d_1 - a_2 b_1 d_2, \nonumber\\
&\tilde{C} = -b_2 c_1 c_2 + a_1 (c_2)^2 + a_2 c_1 d_2 + b_1 c_2 d_2 + 2 b_2 d_1 d_2, \nonumber\\
&\tilde{D} = -c_1 c_2 d_2 + d_1 (d_2)^2, \nonumber
\end{aligned}\right.
\end{equation}
%
where 
\begin{eqnarray}
a_1&=& -\frac{\beta_D}{K_D} \left(1+ \frac{\gamma_D}{\eta_D + \sigma_D}\right)<0,  \nonumber\\
a_2&=& -\frac{\beta_W}{K_W} \left(1+ \frac{\gamma_W}{\eta_W + \sigma_W}\right)<0, \nonumber\\
b_1&=& -\frac{\Psi_D}{K_W} \left(1+ \frac{\gamma_D}{\eta_D + \sigma_D}\right)<0,    \nonumber\\
b_2&=& -\frac{\Psi_W}{K_D} \left(1+ \frac{\gamma_W}{\eta_W + \sigma_W}\right)<0,   \nonumber\\
c_1&=& \beta_D-\gamma_D- \sigma_D \lesseqqgtr 0,   \nonumber\\
c_2&=& \beta_W-\gamma_W- \sigma_W\lesseqqgtr 0, \nonumber\\
d_1&=& \frac{\Psi_D K_D}{K_W}>0,  \nonumber\\
d_2&=& \frac{\Psi_W K_W}{K_D}>0.  \nonumber
\end{eqnarray}

Some of the roots of (\ref{eq6}), which are equal to $I_W^*$, are a component of the biologically relevant (real and positive) equilibria of (\ref{eq5}), known as the endemic equilibria. The roots of any cubic polynomial $\tilde{A}\neq0$ can be solved using the cubic formula, and the number of real and complex roots of any cubic equation can be determined by the sign of discriminant,
\[\Delta = 18 \tilde{A}\tilde{B}\tilde{C}\tilde{D}  - 4\tilde{B}^3\tilde{D}  + \tilde{B}^2\tilde{C}^2 - 4 \tilde{A}\tilde{C}^3 - 27 \tilde{A}^2\tilde{D} ^2.\]
If $\Delta > 0$, then the equation has three distinct real roots. If $\Delta = 0$, then the equation has a multiple root and all of its roots are real. If $\Delta< 0$, then the equation has one real root and two non-real (complex conjugate) roots.

Since the signs of coefficients of (\ref{eq6}) depend on the relative magnitudes of the parameters, we cannot determine any existence criteria about any biologically relevant endemic equilibria. However, in order to examine the impacts that the cultivated area's shape has on interspecies disease transmission,  we can assume that the number of total infections present in a species due to intraspecies interactions is equal to that species' disease recovery rate and mortality rate; that is, let  $c_1=0$, $\beta_D=\gamma_D +\sigma_D$, and $c_2=0$, $\beta_W=\gamma_W +\sigma_W$. This assumption reduces variability of the signs of coefficients of (\ref{eq6}),
\begin{eqnarray}
 \tilde{A} &=& a_1 (a_2)^2 - a_2 b_1 b_2 \lesseqqgtr 0, \nonumber\\
\tilde{B} &=&  (b_2)^2 d_1 - a_2 b_1 d_2\lesseqqgtr0, \nonumber\\
\tilde{C} &=& 2 b_2 d_1 d_2 <0, \nonumber\\
\tilde{D} &=& d_1 (d_2)^2>0. \nonumber
\end{eqnarray}
When $\tilde{A}>0$ or $\tilde{A}<0$, applying Descartes' rules of signs will result in different number of positive, negative, and complex roots of (\ref{eq6}), depending on the sign of the discriminant, as summarized on Table (1) and Table (2).

\begin{table}[!htp]\label{table1}
 \begin{center}
   \caption{The different cases (I)--(IV) describe the number of real and complex roots that the cubic polynomial (\ref{eq6}) has when the coefficient of the cubic term and the discriminant of the cubic formula are positive and/or negative.}
    \begin{tabular}{c|c|c}
     $\tilde{B}>0$ & $\Delta>0$ & $\Delta<0$\\
      \hline
      \multirow{4}{*}{ $\tilde{A}>0$} &\textbf{(I)} & \textbf{(II)}  \\
      & 2 positive & 0 positive \\ 
      & 1 negative & 1 negative\\ 
        & 0 complex & 2 complex\\
      \hline
      \multirow{4}{*}{ $\tilde{A}<0$}  &\textbf{(III)} & \textbf{(IV)}  \\
      & 3 positive & 1 positive\\ 
      & 0 negative & 0 negative\\
       & 0 complex & 2 complex\\ 
    \end{tabular}
      \end{center}
\end{table}

\begin{table}[!htp]\label{table2}
 \begin{center}
   \caption{The different cases (I)--(IV) describe the number of real and complex roots that the cubic polynomial (\ref{eq6}) has when the coefficient of the cubic term and the discriminant of the cubic formula are positive and/or negative.}
    \begin{tabular}{c|c|c}
    $\tilde{B}<0$ & $\Delta>0$ & $\Delta<0$\\
      \hline
      \multirow{4}{*}{ $\tilde{A}>0$} &\textbf{(I)} & \textbf{(II)}  \\
      & 2 positive & 0 positive \\ 
      & 1 negative & 1 negative\\ 
        & 0 complex & 2 complex\\
      \hline
      \multirow{4}{*}{ $\tilde{A}<0$}  &\textbf{(III)} & \textbf{(IV)}  \\
      & 1 positive & 1 positive\\ 
      & 0 negative & 0 negative\\
       & 2 complex & 2 complex\\ 
    \end{tabular}
      \end{center}
\end{table}

We see that when $\beta_D=\gamma_D+\sigma_D$ and $\beta_W=\gamma_W+\sigma_W$, then there is at least one biologically relevant (positive) equilibria of the system, 
the basic reproductive number reduces to
 \[
 \mathcal{R}_0=\frac{\beta_D\beta_W+\sqrt{\beta_D\beta_W\Psi_D \Psi_W}}{\beta_D\beta_W},
 \]
 which is always greater than one. It follows that the disease-free equilibrium is unstable and the system's trajectories tend towards at least one positive endemic equilibrium.

\subsubsection{Numerical Solutions}



To test the epidemiological impact of landscape metrics, we tested the sensitivity of system trajectories to variation in the shape index and contact zone depth. Using the parameters specified in Table (3) we considered the effect of changes in the shape index and contact zone depth on the peak prevalence, the epidemic spread rate, and the endemic equilibrium. We also examined how variation in the cross-species and within-species infections rates impact relative prevalence. We consider the endemic equilibrium more sensitive if a change in an examined parameter results in a larger relative change in the disease prevalence.

\begin{table}[!htp] \begin{center}
\scriptsize
\begin{tabular}{lllll} \toprule
\multicolumn{2}{l}{Variable/ Parameter}              & Base Value                      & Range Value        \\ \midrule
$S_i$ & Susceptible population       & 300, 200 & n/a         \\ 
$I_i$ & Infectious population & 0, 80 & n/a      \\ 
$R_i$ & Recovered population &   0, 20        & n/a         \\
$ \delta_i$ & Cross-species infection rate      &0.3        &      0 -- 1          \\
$ \beta_i $ & Same-species infection rate      &0.3        &      0 -- 1         \\
$\gamma_i$ & Recovery rate      &0.1        &      n/a          \\
$k$ & Scaling constant       &      3.55     & n/a       \\
$d$ & Penetration distance      & 0.5 & 0 --  0.5296       \\
$\Lambda_i$ & Per capita birth rate \!\! & 30 & n/a\!\!   \\
$\ell$ & Perimeter length &     16       &16 -- 1000   & \\
$a$ & Domesticated habitat area              &     500         & n/a & \\
$z$ & Total area       &      1000      & n/a         \\
$\eta_i$	  & Loss of immunity rate             &  0.3                          & n/a       \\
$\sigma_i$	  & Death rate           &  0.1                          & n/a        \\
$\mu$	  & Shape index             &  1, 12                         & 1 -- 12           \\ \bottomrule
\end{tabular}
\caption{Parameter definitions, units, and values for the model with $i=D,W$.\\}
\end{center} 
\normalsize
\end{table}

As a baseline, we used initial conditions for the wild population, assuming no interaction between the wild and domesticated populations, (isolated populations) $\Psi_W = 0$, and solved for the endemic equilibrium of the wild population ($S_W, I_W, R_W$). We then varied the domesticated population parameters only, in order to explore the general epidemiological effects of the introduction of a non-native disease-free population into an area where disease is endemic to the wild population. The results are shown in the following simulations.





\begin{figure}[h]
\centering
\subfloat[]{{\includegraphics[scale=.115]{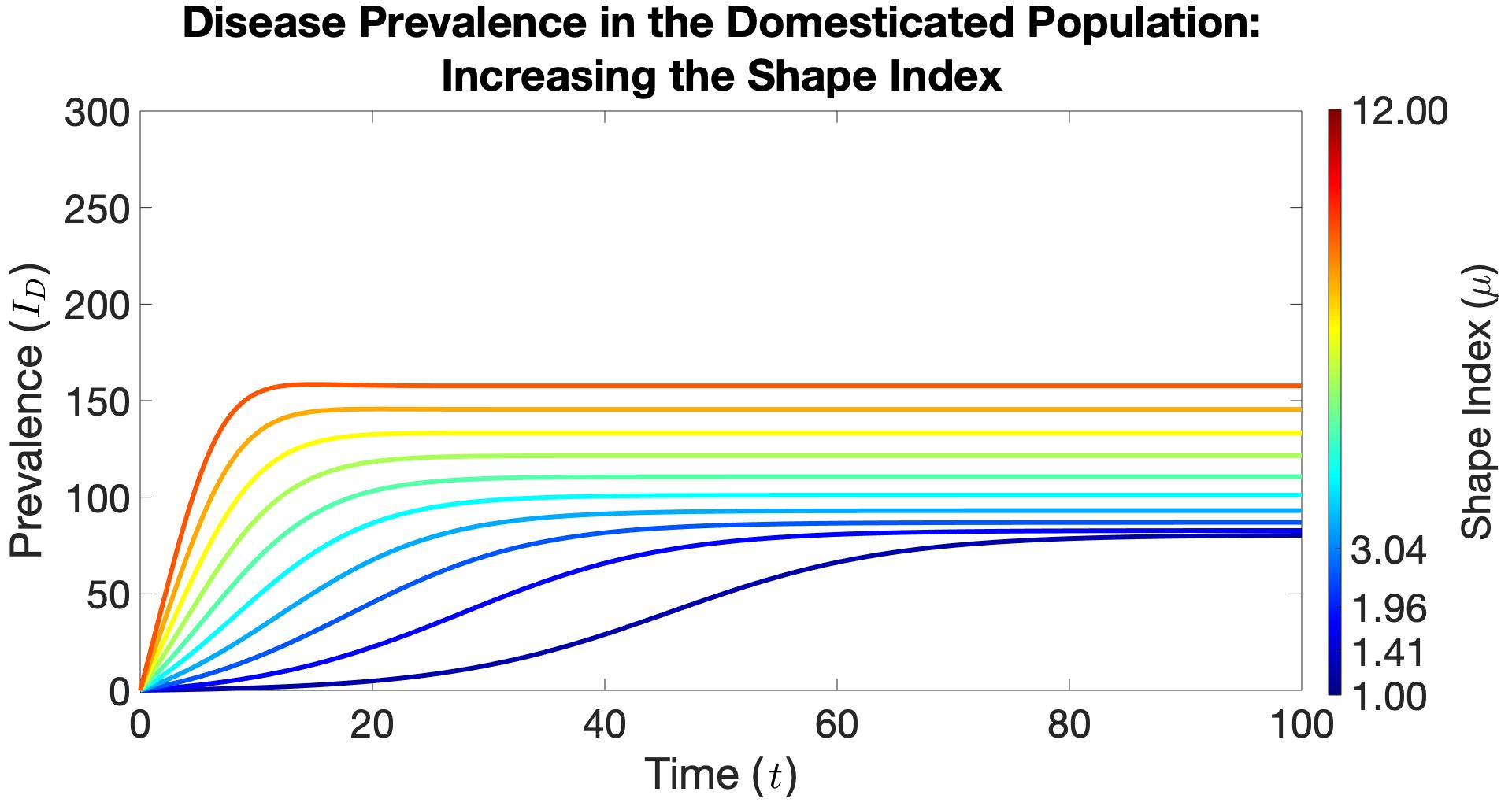}\label{fig:2popMuDom}}}

\subfloat[]{{\includegraphics[scale=.115]{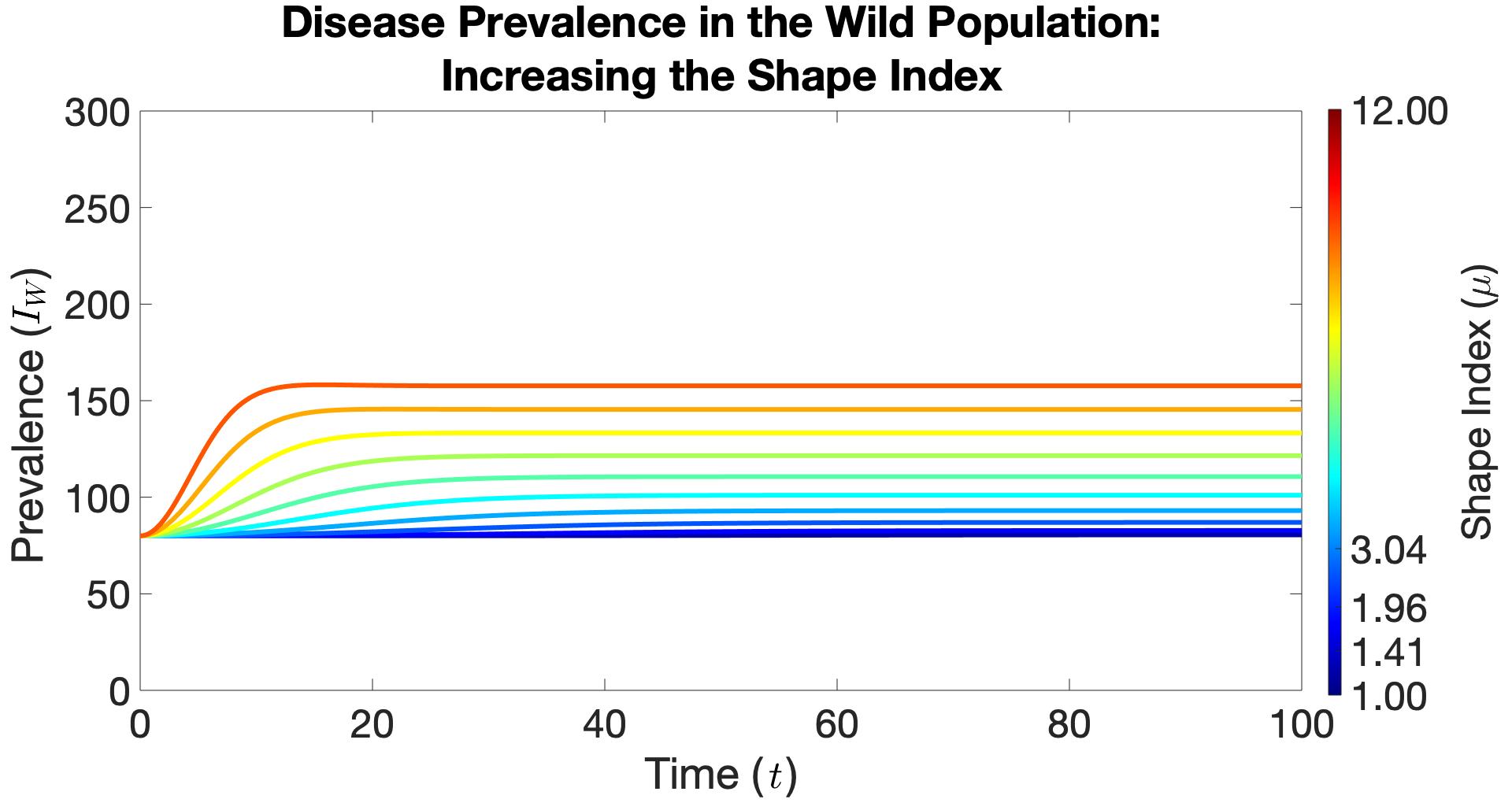}\label{fig:2popMuWild}}}

\caption{(a) The domesticated population's prevalence as the shape index, $\mu$, increases from a minimum value of $\mu=1$ (the darkest blue line) to an arbitrarily assigned value of $\mu=12$ (the darkest red line). 
(b) The wild population's prevalence as the shape index, $\mu$, increases from a minimum value of $\mu=1$ (the darkest blue line) to an arbitrarily assigned value of $\mu=12$ (the darkest red line). $\mu=1.00$ corresponds to the shape index of a circle, $\mu=1.41$ corresponds to the shape index of a rectangle (4 units long and 1 unit wide),  $\mu=1.96$ corresponds to the shape index of a rectangle (10 units long and 1 unit wide), and $\mu=3.04$ corresponds to the shape index of a fernleaf.}
\label{fig:de}
\end{figure}

In Figures (\ref{fig:2popMuDom}) and (\ref{fig:2popMuWild}), the converted regions' area $a$ is held constant while the perimeter $\ell$ is increased, which changes the shape index from that of a circle, $\mu=1$ to that of an irregular shape, $\mu=12$. The contact zone depth is assumed to be constant. Note that with an increase in the converted area's shape index, implying a higher amount of habitat fragmentation over a landscape, there is an increase in peak prevalence, the rate of epidemic spread, and the endemic equilibrium within the domesticated and wild populations. The more compact is the converted area, the lower is the rate of disease spread.




\begin{figure}[h]
\centering
\subfloat[]{{\includegraphics[scale=.115]{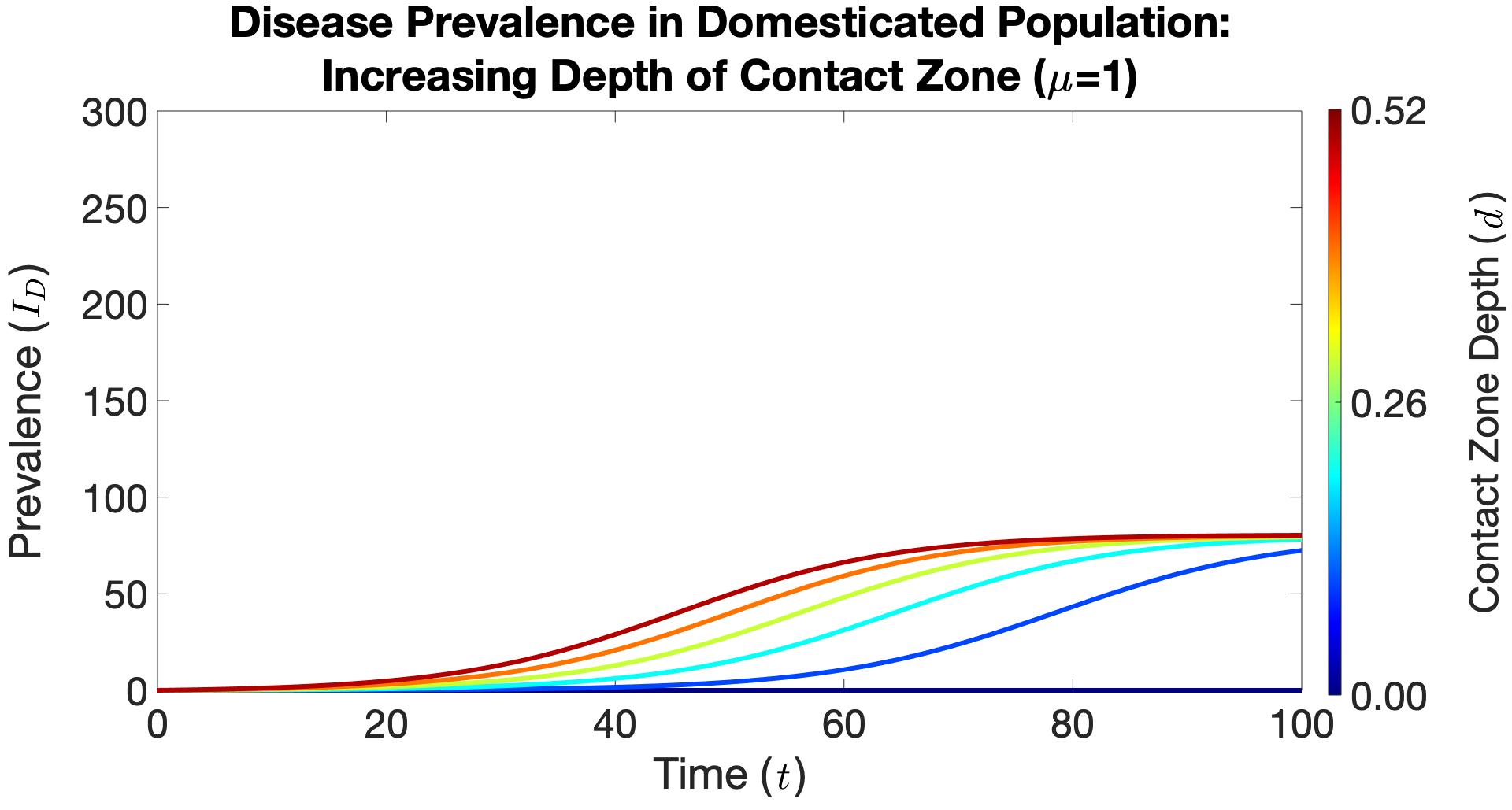}\label{fig:2popDepthDom1}   }}

\subfloat[]{{\includegraphics[scale=.115]{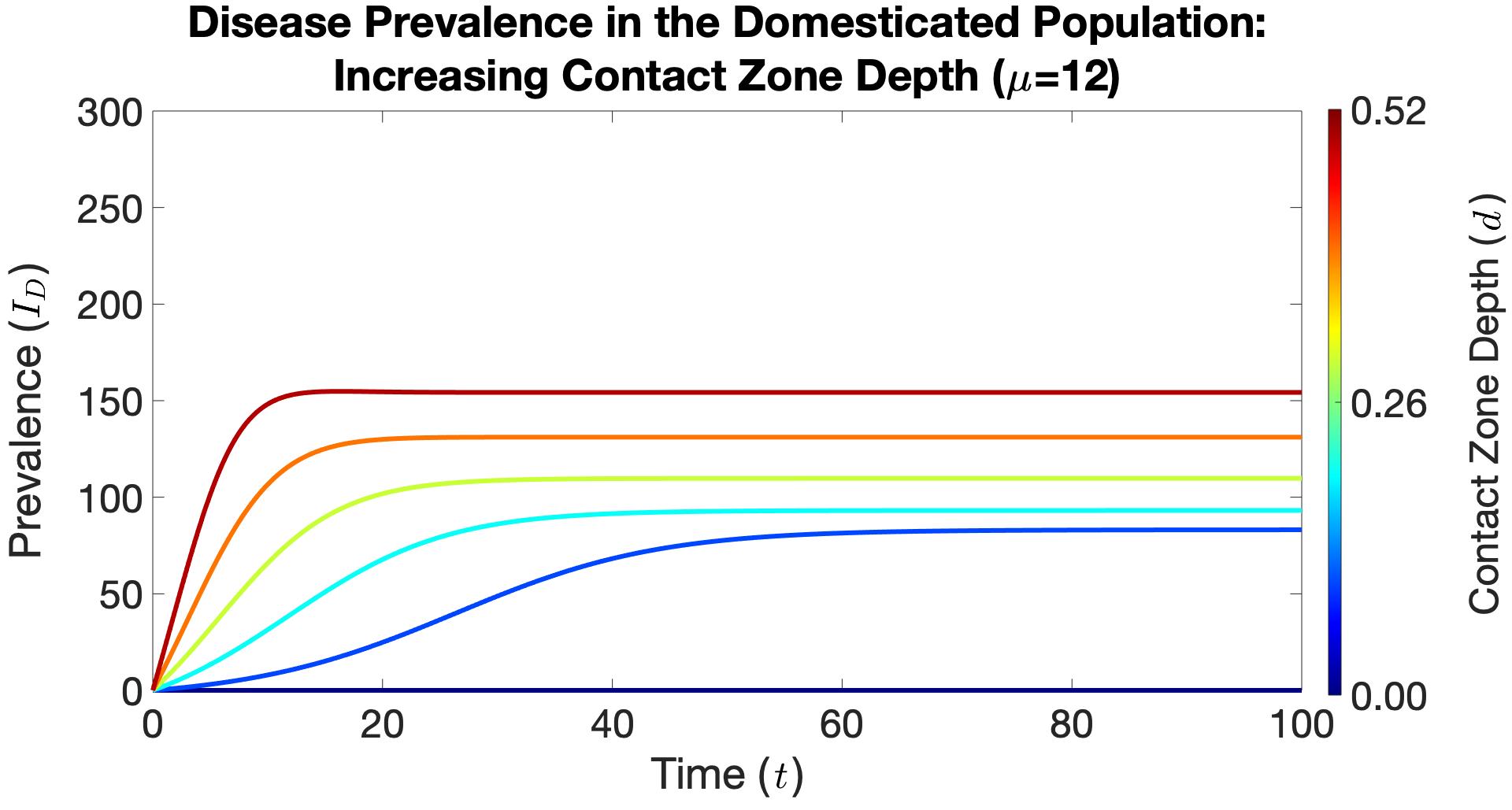}\label{fig:2popDepthDom12}  }}

\caption{Domesticated Species. (a) As the contact zone depth, $d$, in a compact converted area increases from zero (represented in dark blue) to an arbitrarily large maximum (represented in red), there is a slight increase in peak prevalence, the rate of epidemic spread, and the endemic equilibrium.
(b) As the contact zone depth, $d$, in a converted area with an irregular shape, $\mu=12$, increases, there is a significantly greater increase in peak prevalence, rate of epidemic spread, and endemic equilibrium.
}
\label{fig:de}
\end{figure}



\begin{figure}[h]
\centering
\subfloat[]{{\includegraphics[scale=.115]{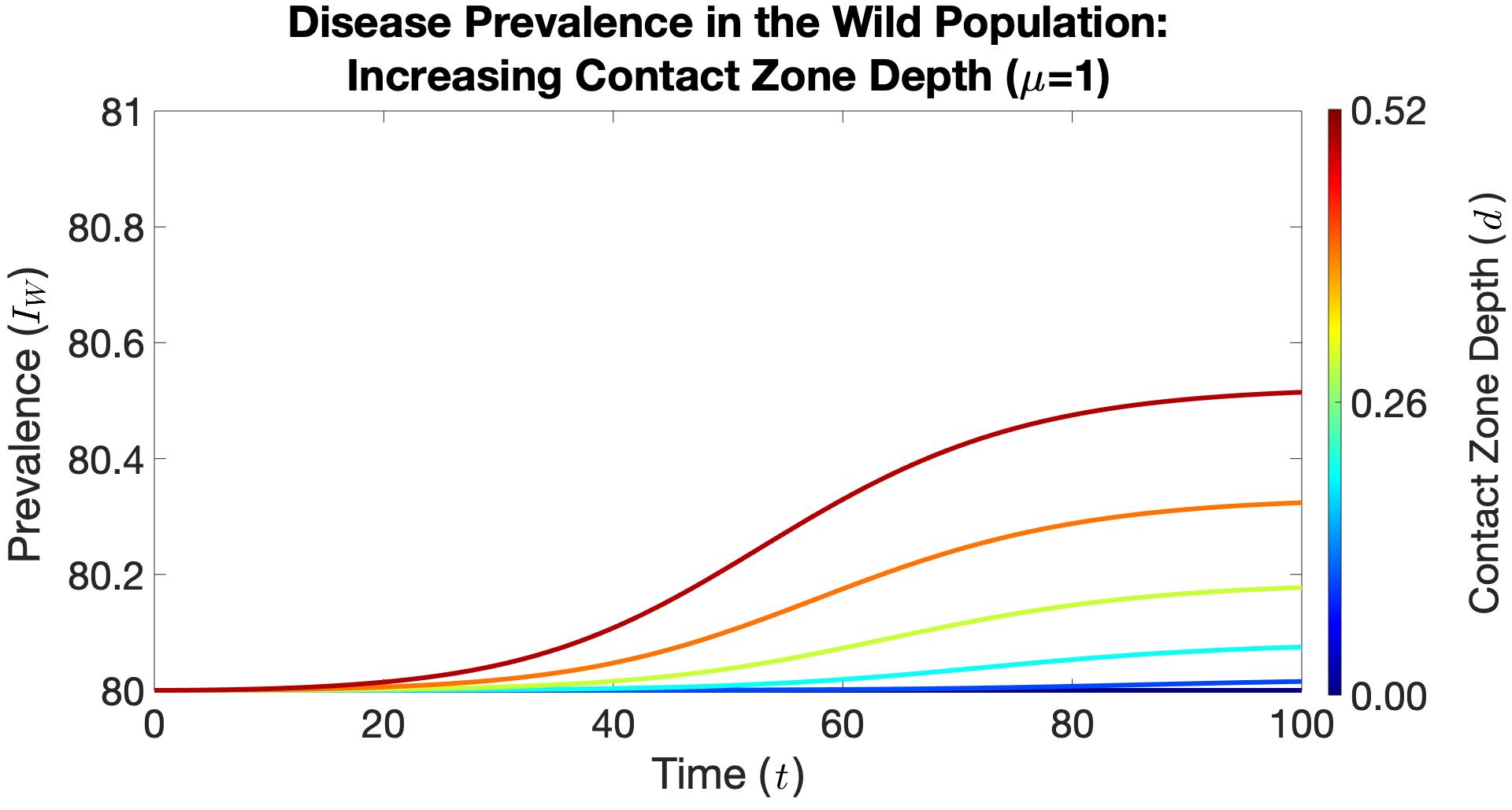}\label{fig:2popDepthWild1}}}

\subfloat[]{{\includegraphics[scale=.115]{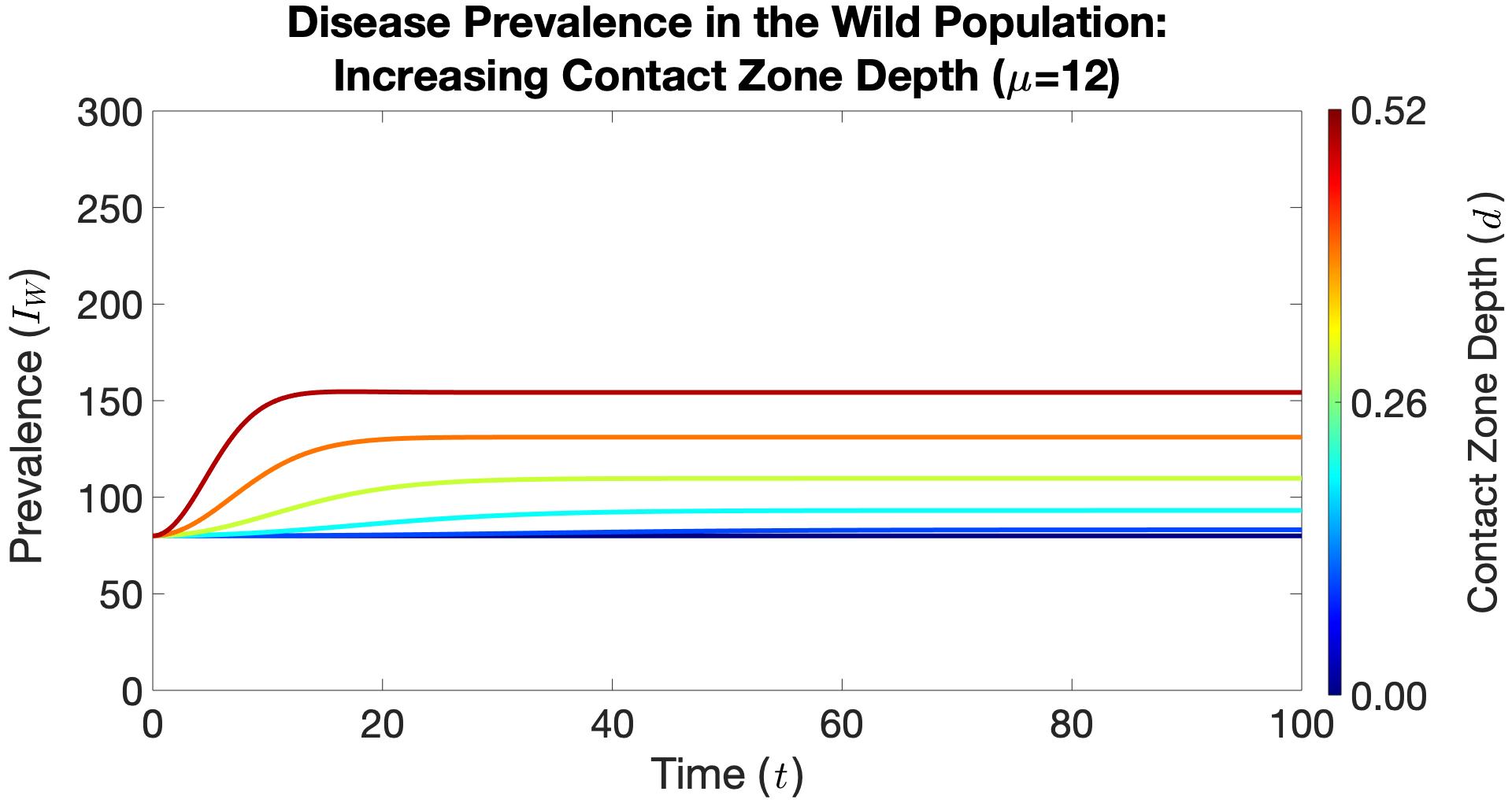}\label{fig:2popDepthWild12}}}

\caption{Wild Species. (a) As the contact zone depth, $d$, in a compact converted area increases from zero  (represented in dark blue) to an arbitrarily large maximum (represented in red), there is an increase in peak prevalence, rate of epidemic spread, and endemic equilibrium.
(b) As the contact zone depth, $d$, in a compact converted area with a shape index of $\mu=12$ (some irregular shape) increases, there is a much larger increase (compared to the case of $\mu=1$) in peak prevalence, the rate of epidemic spread, and endemic equilibrium.
}
\label{fig:de}
\end{figure}

Figures (\ref{fig:2popDepthDom1}) and (\ref{fig:2popDepthWild1}), illustrate the effect that an increase in the contact zone depth, $d$, has on disease prevalence, assuming that the domesticated population inhabits a circular converted region. Figures (\ref{fig:2popDepthDom12}) and (\ref{fig:2popDepthWild12}) show similar results, but the domesticated population is assumed to inhabit an irregularly shaped area. In all cases, it is assumed that initially there is no habitat overlap between the wild and domesticated populations. As the contact zone depth between the wild and domesticated species increases, there is an increased peak prevalence, and faster rate of disease spread. Also, the less compact the converted area, the greater the impact on disease prevalence and the rate of epidemic spread.





\begin{figure}[h]
\centering
\subfloat[]{{\includegraphics[scale=.115]{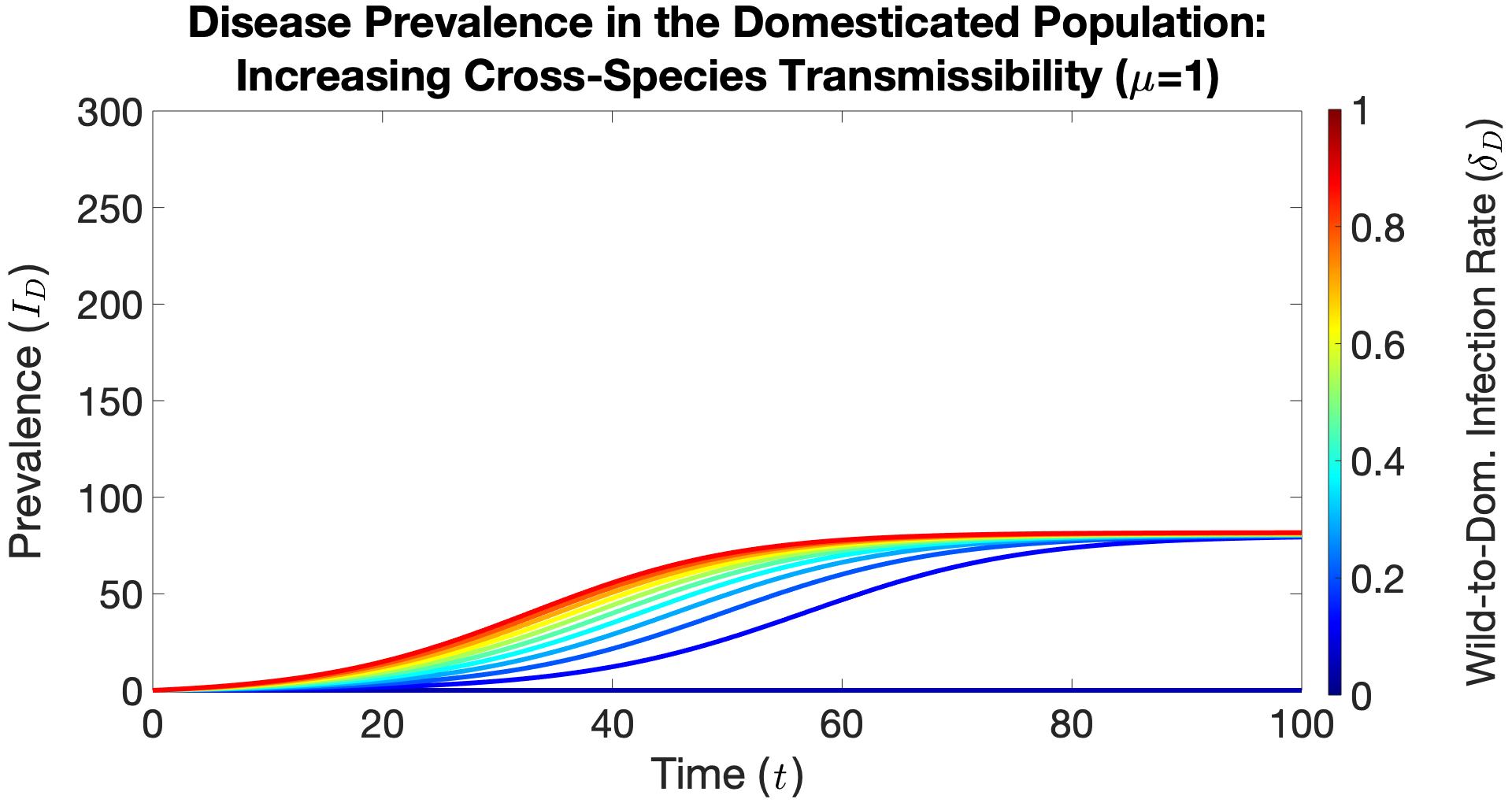}\label{fig:2popDeltaDom1} }}

\subfloat[]{{\includegraphics[scale=.115]{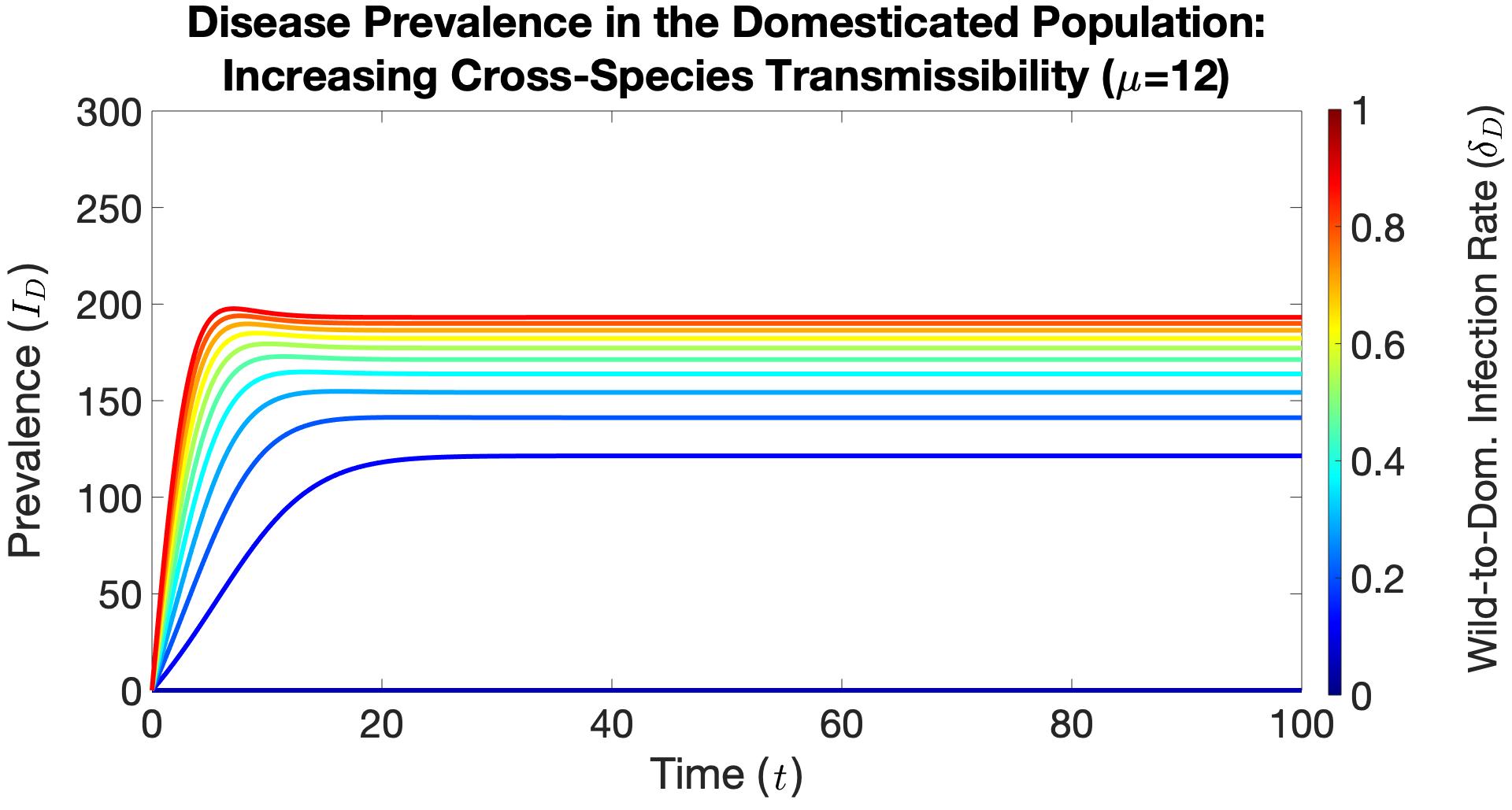}\label{fig:2popDeltaDom12} }}

\caption{Domesticated Species. (a)  As the cross-species infection rate increases from $\delta_D=0.1$ (the darkest blue line) to $\delta_D=1.0$ (the darkest red line), for a converted area in the shape of a circle, $\mu=1$, there is a modest increase in peak prevalence and rate of epidemic spread in the domesticated population.
(b) For a converted area with a shape index of $\mu=12$, as the cross-species infection rate increases from $\delta_D=0.1$ to $\delta_D=1.0$, there is a significantly greater increase in peak prevalence, the rate of epidemic spread, and then endemic equilibrium in the domesticated population.
}



\end{figure}


\begin{figure}[h]
\centering
\subfloat[]{{\includegraphics[scale=.115]{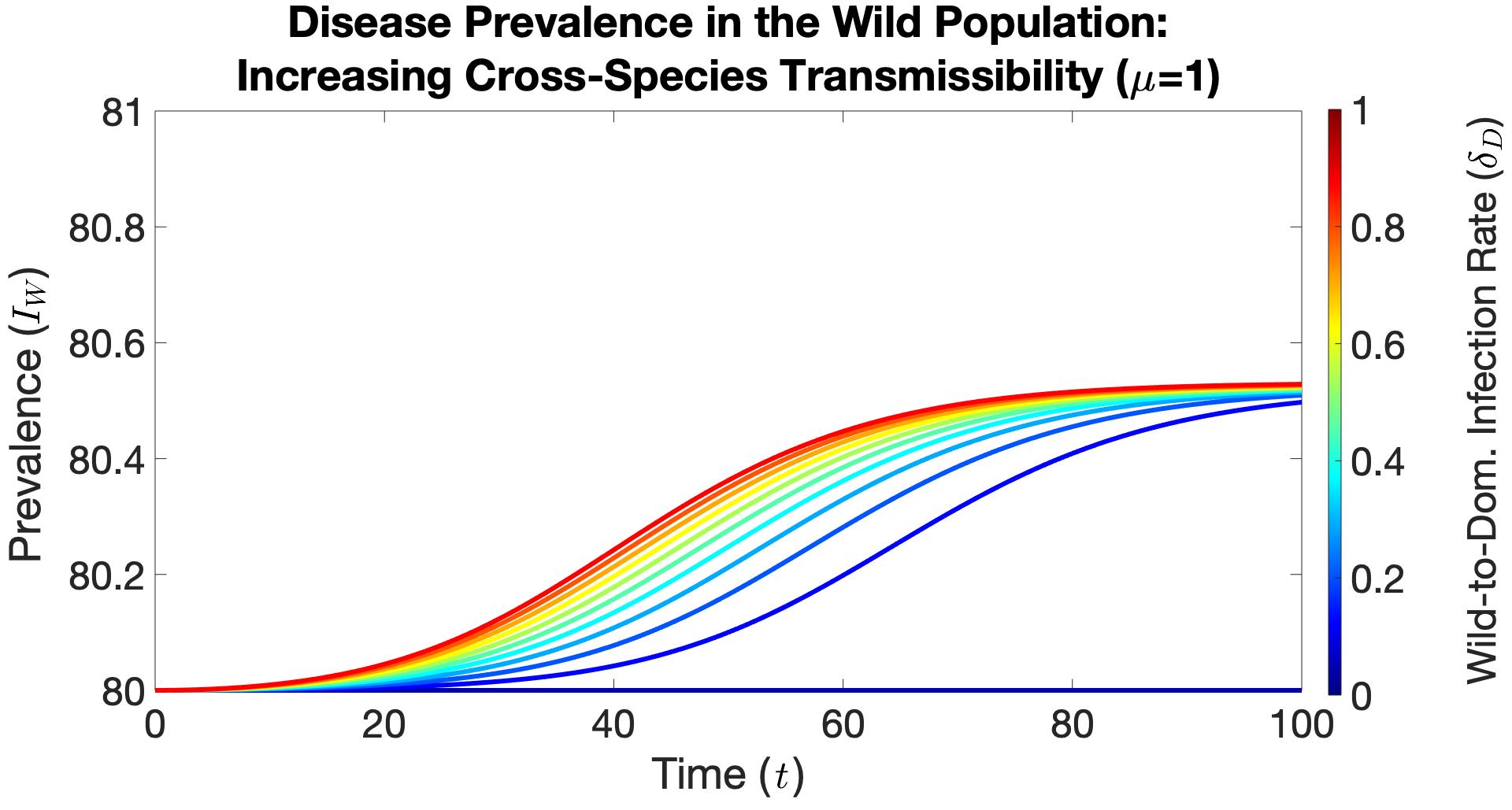}\label{fig:2popDeltaWild1}}}

\subfloat[]{{\includegraphics[scale=.115]{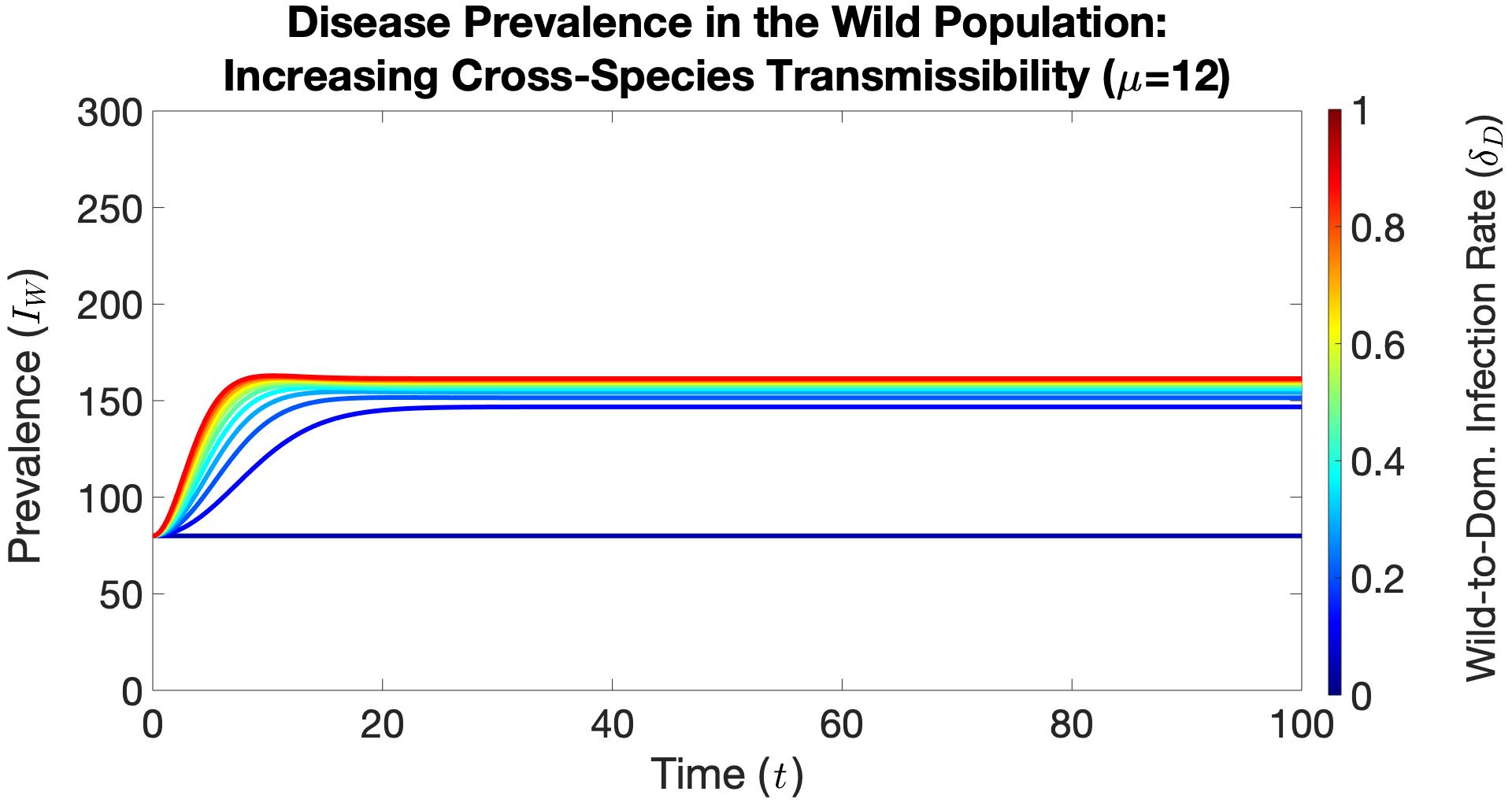}\label{fig:2popDeltaWild12}}}

\caption{Wild Species. (a)  As the cross-species infection rate increases from $\delta_D=0.1$ (the darkest blue line) to $\delta_D=1.0$ (the darkest red line), for a converted area in the shape of a circle, $\mu=1$, there is a very small increase in peak prevalence and the rate of spread of the epidemic in the wild population.
(b) For a converted area with a shape index of $\mu=12$, as the cross-species infection rate increases from $\delta_D=0.1$ to $\delta_D=1.0$, there is a large increase in peak prevalence, the rate of epidemic spread, and the endemic equilibrium in the wild population.
When comparing \ref{fig:2popDeltaWild1}  to \ref{fig:2popDeltaWild12}, for a particular value of the wild-to-domesticated infection rate, there is a higher level of disease prevalence in Figure \ref{fig:2popDeltaWild12}.} 

\end{figure}



\begin{figure}[h]
\centering
\subfloat[]{{\includegraphics[scale=.115]{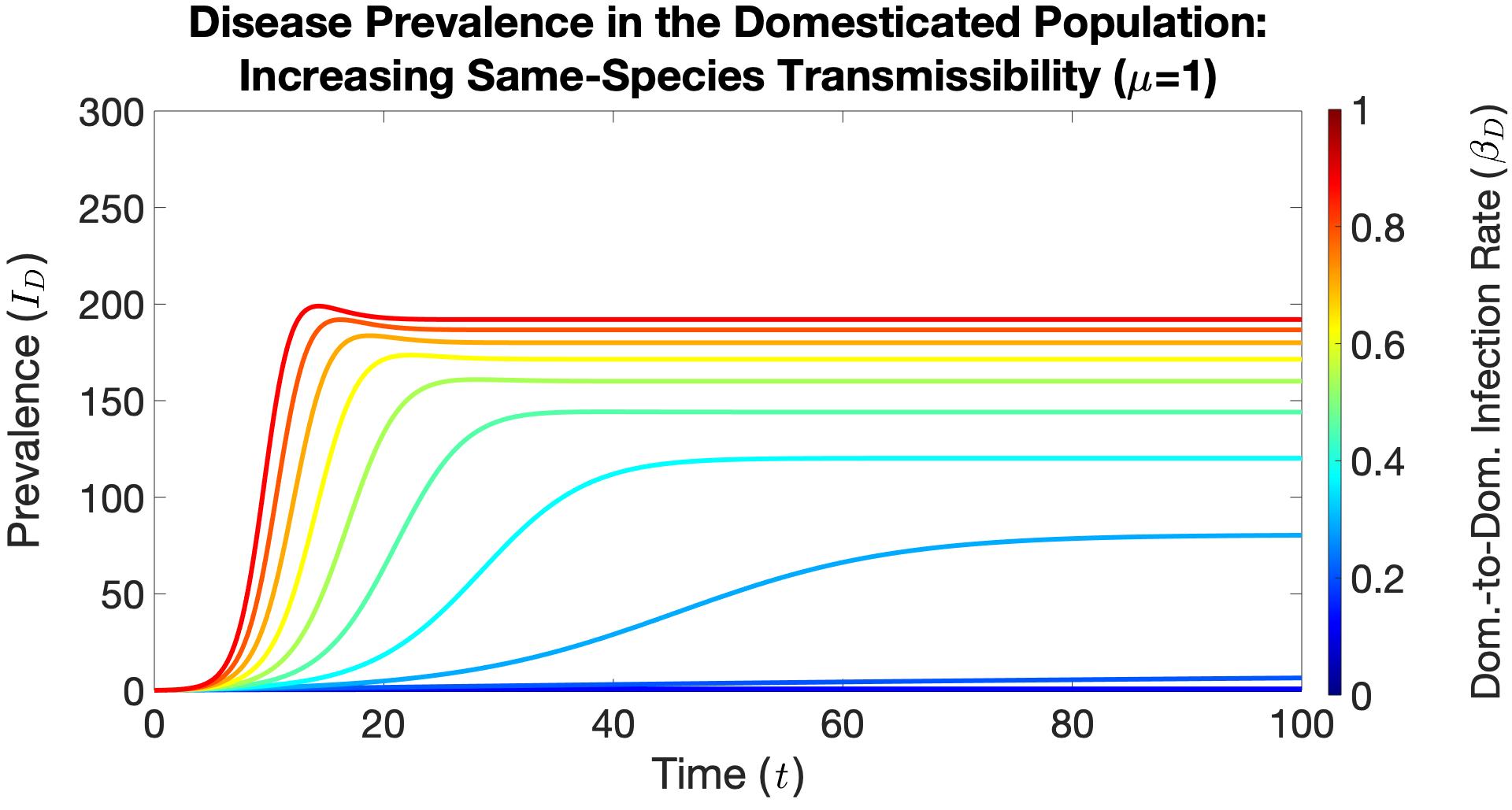}\label{fig:2popBetaDom1}}}

\subfloat[]{{\includegraphics[scale=.115]{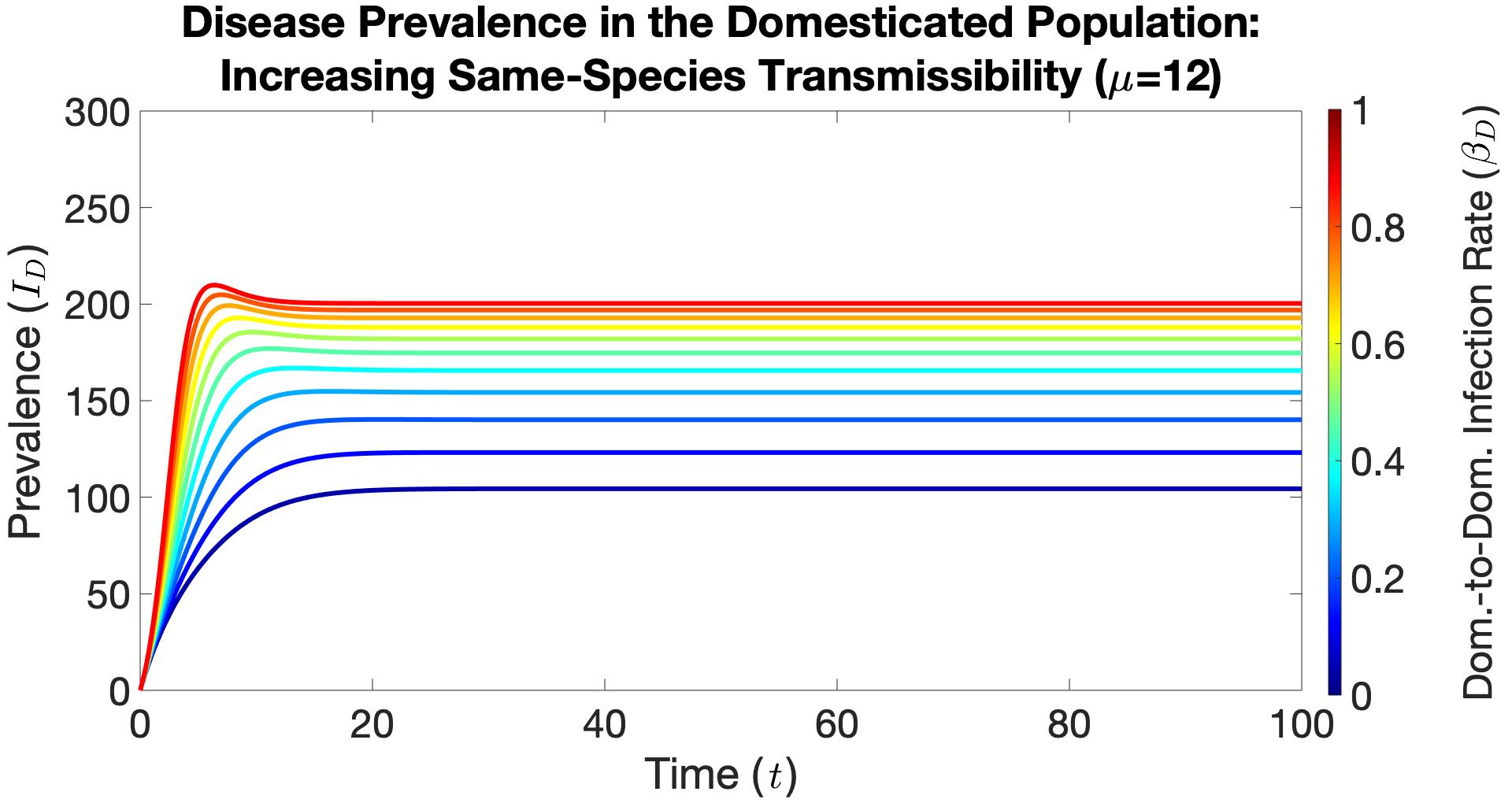}\label{fig:2popBetaDom12}}}

\caption{Domesticated Species. (a) On a circular converted area, $\mu=1$, as the within-species transmission rate among domesticated animals increases from $\beta_D=0$ (the darkest blue line) to $\beta_D=1$ (the darkest red line), there is an increase in the peak prevalence, the rate of epidemic spread, and the endemic equilibrium among the domesticated animals. 
(b) As the transmission rate increases from $\beta_D=0$ to $\beta_D=1$ on an irregularly shaped landscape, $\mu=12$, there is an increase in the peak prevalence, rate of epidemic spread, and endemic equilibrium among domesticated animals, but at higher levels than if the converted area was a circle.}
\label{fig:de}
\end{figure}

\begin{figure}[h]
\centering
\subfloat[]{{\includegraphics[scale=.115]{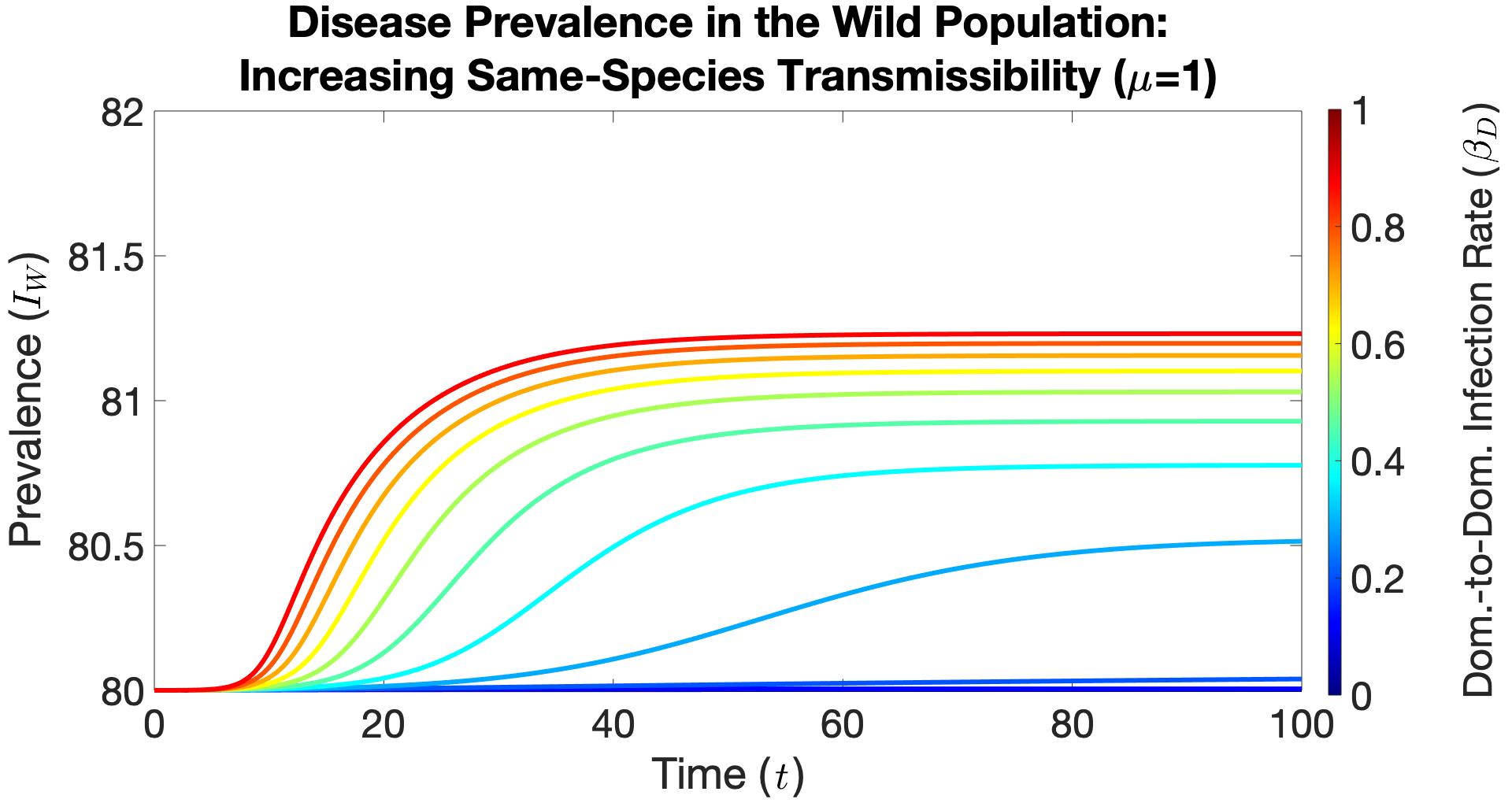}\label{fig:2popBetaWild1}}}

\subfloat[]{{\includegraphics[scale=.115]{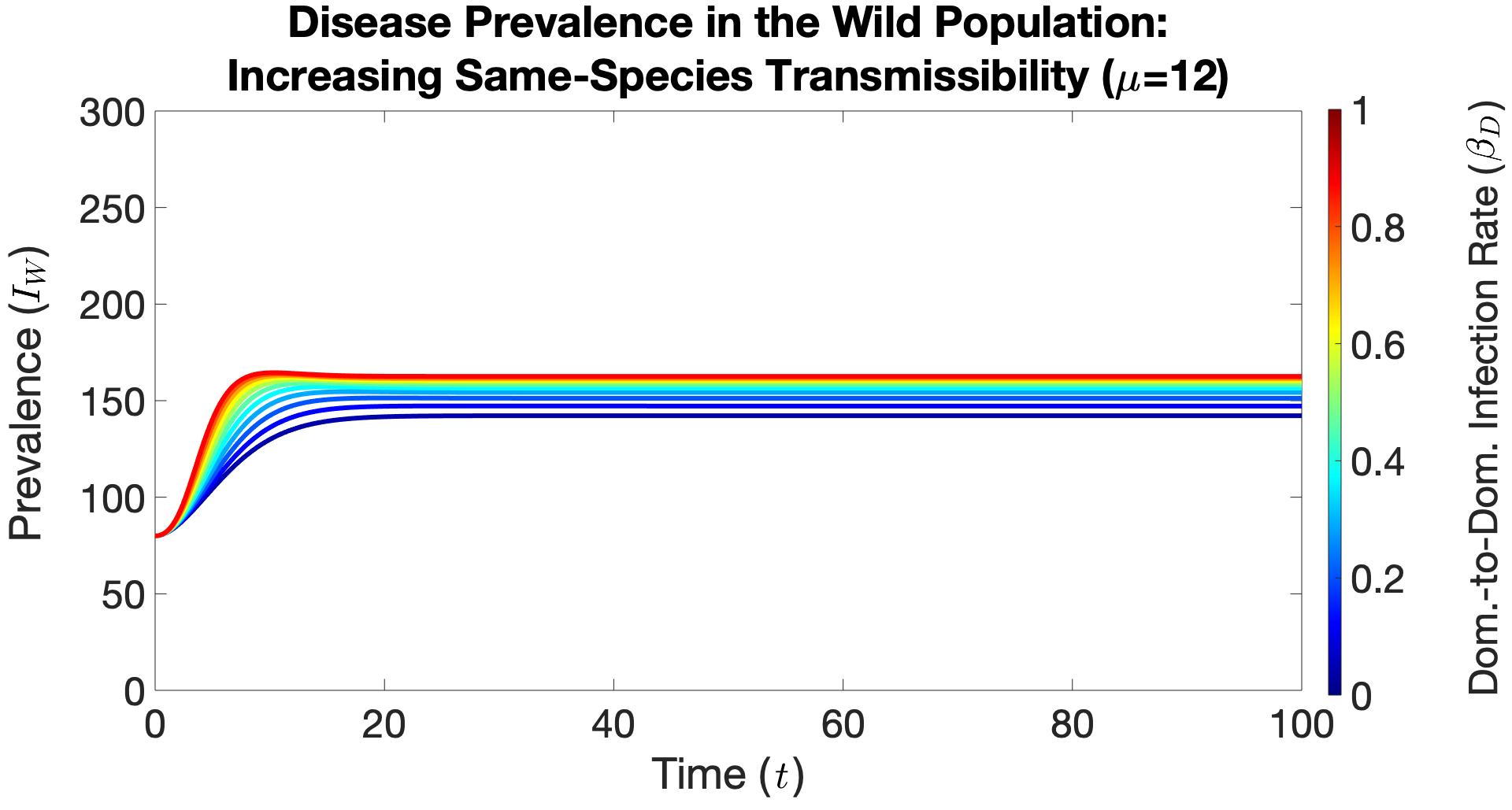}\label{fig:2popBetaWild12}}}

\caption{Wild Species. (a) On a circular converted area, $\mu=1$, as the within-species transmission rate among domesticated animals increases from $\beta_D=0$ (the darkest blue line) to $\beta_D=1$ (the darkest red line), there is an increase in the the peak prevalence, rate of epidemic spread, and endemic equilibrium among the wild population. 
(b) As the same transmission rate increases from $\beta_D=0$ to $\beta_D=1$ on an irregularly shaped landscape, $\mu=12$, there is an increase in the peak prevalence, rate of epidemic spread, and endemic equilibrium wild animals, but, in general, at higher levels than if the cultivated area was a circle. The impact on wild animals is much smaller than on domesticated animals.}
\label{fig:de}
\end{figure}

Figures (\ref{fig:2popDeltaDom1}) -- (\ref{fig:2popBetaWild12}) illustrate the effect that the shape of the converted area has on the impact of changes in cross-species (Figures (\ref{fig:2popDeltaDom1}) -- (\ref{fig:2popDeltaWild12})) and within-species (Figures (\ref{fig:2popBetaDom1}) -- (\ref{fig:2popBetaWild12})) transmission rates. All show that the more irregular the shape of the converted area, the faster an epidemic spreads, and the greater the long-run impact an epizootic infection has on both wild and domesticated species. There are, however, significant differences in the effects of changes in cross-species and within-species transmission rates.





Figures (\ref{fig:2popDeltaDom1}) -- (\ref{fig:2popDeltaWild12}) show that the effect of an increase in cross-species transmission rates saturates more rapidly in less compact, $\mu=1$, converted areas. That is, for a compact converted area, the endemic equilibria are relatively insensitive to further increases in cross-species transmission rates, and where converted areas are less compact, the endemic equilibria continue to increase with increasing transmission rates.





Figures (\ref{fig:2popBetaDom1}) -- (\ref{fig:2popBetaWild12}) show that the endemic equilibria exhibit lower prevalence in more compact converted areas than in less compact (irregularly shaped) converted areas, and the endemic equilibria are relatively sensitive for both compact and less compact converted areas.
Furthermore, Figures (\ref{fig:2popBetaDom1}) -- (\ref{fig:2popBetaWild12}) show that as the shape index increases, so do the epidemiological effects of increasing transmission rates among domesticated animals. Disease prevalence, the rate of epidemic spread, and the endemic equilibrium increase for both the domesticated and wild populations. This is due to an increase in homogenous mixing of the two species, which occurs when the shape index is increased. The impact on domesticated species is, as one would expect, much greater than the on wild species. The rate of spread is faster, and peak prevalence is higher. Nonetheless, a higher edge-to-area ratio of converted land implies larger impacts on disease prevalence among wild species.

In Figures (\ref{fig:2popBetaDom1}), (\ref{fig:2popBetaDom12}), (\ref{fig:2popBetaWild1}), and (\ref{fig:2popBetaWild12}) as the shape index increases so does disease prevalence, the rate of epidemic spread, and the endemic equilibrium for both the domesticated and wild populations. The endemic equilibrium in Figure (\ref{fig:2popBetaDom1}) compared to Figure (\ref{fig:2popBetaDom12}) is, in general, higher for lower levels of local infection. This is due to an increase in homogenous mixing of the two species which occurs when the shape index is increased, and which results in more domesticated infections. This implies that a higher edge-to-area ratio of a converted area, i.e. more habitat fragmentation, can impact the disease prevalence of a population inhabiting that region, even with low disease virulence. Moreover, as the domesticated-to-domesticated infection rate increases, the solutions are stratified more uniformly on an irregular (e.g. $\mu=12$) than on regular (e.g $\mu=1$ ) landscapes. The practical significance of this is that an intervention to  to decrease cross-species infection will have both a greater and a more durable impact in more fragmented than in less fragmented landscapes. In fact, for the most compact converted area, long run disease prevalence is insensitive to changes in the domesticated-to-domesticated infection rate.

For both Figures (\ref{fig:2popBetaDom1}) and (\ref{fig:2popBetaDom12}), a high domesticated-to-domesticated infection rate yields a peak prevalence higher than the endemic equilibrium, which has the same value for both cases. Moreover, with a shape index of $\mu=1$, the disease prevalence is lower for small values of $\beta_D$, implying that the compactness of landscape configuration can reduce the rate of epidemic spread and endemic level of prevalence. On Figures (\ref{fig:2popBetaDom1}) and (\ref{fig:2popBetaDom12}), the peak prevalence is greater than the endemic equilibrium when the domesticated-to-domesticated infection rate $\beta_D$ is very large relative to the recovery rate $\gamma_D$ and loss of immunity rate $\eta_D$. On the other hand, when the domesticated-to-domesticated infection rate $\beta_D$ becomes smaller than the recovery rate $\gamma_D$ and the loss of immunity rate $\eta_D$, the peak prevalence is endemic equilibrium. The physical significance is that more domesticated animals are becoming infected as the same-species infection rate $\beta_D$ increases.

A comparison between Figures (\ref{fig:2popBetaWild1}) and (\ref{fig:2popBetaWild12}), shows that for the most compact converted area (a circle), as the domesticated-to-domesticated infection rate increases from $\beta_D=0$ to $\beta_D=1$ so does the peak prevalence, the rate of epidemic spread, and endemic equilibrium, but at a very small increase relative to the total population. As the shape of the converted area is deformed to $\mu=12$, an increase in the domesticated-to-domesticated infection rate yields a greater increase in peak prevalence, the rate of epidemic spread, and the endemic equilibrium than if the cultivated area was a circle. That is, the incidence of disease is increasing in the degree of landscape fragmentation, even when considering a zoonotic infection that primarily impacts a different species that resides on a connected but separate habitat.

All of the simulations in this manuscript assume that the total population is relatively constant. At the initial time, the number of susceptible domesticated animals $S_D$ is large, with a fixed positive cross-species infection rate $\delta_D$. The peak prevalence is greater than the endemic equilibrium if the infection rate $\delta_D$ or $\beta_D$ are significantly greater than the recovery rate $\gamma_D$ and the loss of immunity rate $\eta_D$, for the domesticated species. When a transmission rate is very large, and the recovery rate and the loss of immunity rate are relatively small, the susceptible population decreases rapidly. The infectious population increases until there are no new susceptible individuals can be infected. The infectious population then decreases until an equilibrium because the recovered domesticated animals lose immunity and are again susceptible to infection.



\section{Discussion}



In this paper, we model the effect of land conversion for livestock production on epizootic infectious disease emergence and transmission. We focus on the effect of patterns of land conversion on the epidemiological edge between converted and unconverted land, or the contact zone within which wild and domesticated species mix. We show how changes in the pattern of land conversion induce changes in pathogen transmission both within fragments and across the landscape \citep{allan2003effect,patz2000effects,service1991agricultural}.  Specifically, we combine concepts from landscape ecology, spatial epidemiology, and mathematical epidemiology to analyze the disease implications of the shape of converted land, obtaining general results on the impact of the size and shape of the edge and its relation to interactions between wild and domesticated populations. We show the relation of zoonotic disease transmission to the edge between wild and domesticated habitats.


Our numerical experiments show that increased edge effects result in more disease prevalence in an introduced domesticated species, at least up to an upper-bound on the shape index. The size of the epidemiological edge depends on two magnitudes: the length of the perimeter of the converted area (the shape index) and the contact zone depth. The contact zone depth depends on the characteristics of the species in both converted and wild habitats. It is also constrained by the shape of the converted area. A converted area in the shape of a circle ($\mu =1$) allows for maximum contact zone depth. For all other shapes ($\mu >1$), if the contact zone is of uniform depth throughout the perimeter, then as the area of the core habitat goes to zero, some portion of the converted area will lie outside of the contact zone. 

We show that an increase in the depth of the contact zone in a more compact converted area results in less disease prevalence as compared to the same increase in contact zone depth in a less compact converted area. Increasing either the shape index or contact zone depth also increases the size of the epidemiological edge, and hence disease prevalence. This holds for all models. Both the basic reproductive number and the endemic equilibrium also increase with increases in the shape index or the contact zone depth. These results suggest that more minimizing the habitat edge can buffer epizootic disease transmission.

We also find that with homogeneous mixing, an increase in the size of the contact zone relative to the size of the converted area increases the size of the susceptible domesticated population in potential contact with the infected wild population. Since the size of the contact zone depends in part on the shape of the converted area, this implies that the disease risks to the susceptible population may be regulated, at least in part, by managing the shape of the converted area. If the impacts of disease transmission are positive, it will be desirable to increase the size of the contact zone. If the impacts are negative it will be desirable to reduce it.

Our results support the findings of \citep{white2018disease} and a number of other phenomenological models. That is, increased edge effects lead to higher pathogen transmission and ultimately pathogen persistence. They differ from the findings of \citet{faust2018pathogen}, who reported that intermediate levels of habitat fragmentation and permeability contributed to the highest levels of disease prevalence and persistence. The difference arises in part because of differences in model assumptions. \citet{faust2018pathogen} did not consider the shape of the converted area, which mediates zoonotic pathogen transmission, and assumed that domesticated and wild population growth depends logistically on habitat availability for the respective species. We do consider the shape of the converted area, but assume a constant birth rate. This potential limitation of our model can be improved by allowing population growth to depend on the size of habitat available. Nonetheless, the current model is suitable as a general theoretical framework.

Epidemiological models have generated inconsistent results on the impact of spatial heterogeneity on pathogen transmission and persistence. There are different answers to the ecological-epidemiological question: does more habitat fragmentation promote or impede disease transmission and persistence? If fragmentation involves the creation of dispersed and unconnected islands of natural habitat, it can lead to the isolation of sub-populations of wild species. If there is no disease transmission between wild and domesticated species this can reduce disease transmission. On the other hand, if fragmentation increases the size of the epidemiological edge, it can increase disease transmission between domesticated and wild species. Our model does not consider the first of these. But it does have implications for the management of disease through patterns of conversion.  Where the wild species in an area may be a source of infection to domesticated species, or vice versa, disease transmission may be reduced by reducing the size of the epidemiological edge, whether through the shape of the converted area or the contact zone depth.

%
%
%


There is an important empirical literature on the relation between land conversion, habitat fragmentation, biodiversity and zoonotic disease emergence and transmission. This literature finds that isolated small patches exhibit both low biodiversity and high disease incidence \citep{logiudice2003ecology, allan2003effect}. Instead we focus on spread as a function of contact between susceptible and infected individuals within contact zones at the margins between converted and unconverted land. We consider how the shape and size of converted land relative to unconverted land affects the peak prevalence, endemic prevalence, and the rate at which zoonotic disease spreads.

Our study provides insight about how the epidemiological effects of the introduction of a disease free-population into an area where zoonoses are endemic depends on the pattern of land conversion. Patterns of land conversion that generate greater edge effects will, other things being equal, generate higher disease risks. The framework described here is potentially applicable to a number of emerging and re-emerging zoonotic and epizootic diseases. Notably, vector-borne diseases, such as malaria or dengue, have the highest incidence at the human-wildlife interface \citep{barua2013hidden,vanwambeke2007impact}. The incorporation of land conversion patterns on vector distribution and serotype dominance throughout a landscape can provide insight into effective vaccination strategies and mosquito eradication campaigns that optimally reduce disease transmission while reducing costs \citep{kabir2020cost, tennakone2018host}.

%
%
%
%
%
%

Another (non-vector) example is rabies, which is found in wild species such as bats, skunks, raccoons, and foxes, and can be transmitted to domesticated species such as cattle and dogs. Most mammals capable of contracting rabies exhibit short latent periods, and the disease is almost always fatal, so the vital dynamics proposed are appropriate. Moreover, expansion of the converted area for the purposes of livestock production increases the size of the contact zone, and hence the likelihood of contact with rabies-infected wild species. The model can further be extended to a metapopulation model by incorporating multiple wild populations. Pursuit of this model will allow for a more robust understanding of how the incorporation of the edge impacts the spread of zoonotic diseases.

Land conversion in tropical forests has been implicated in the emergence of many novel zoonotic diseases \citep{jones2008global}. The origin of most emerging and re-emerging zoonoses are fragmented forest margins in Central and South America, Sub-Saharan Africa, and South Asia, where people have limited capacity to manage the sanitary risks posed by infected wild species, and where land conversion via deforestation is common \citep{allan2003effect, service1991agricultural, crooks2011global}. The disease implications of habitat fragmentation and the ensuing edge effects, primarily cross-species pathogen transmission, is a reason for concern about deforestation. While the disease implications of deforestation would be expected to saturate, the risks are currently still increasing.  At present, about 70\% of the world's remaining forest cover lies within 1km of the forest edge \citep{haddad2015habitat}. Continued population growth in the poorest regions of the world increases the impetus for land conversion, and the ecological and epidemiological consequences that follow. In fact, outbreak risks are highest in regions where population-driven expansion into wildlife refugia brings susceptible people or livestock into contact with wildlife reservoirs of diseases. This has created a demand for models that capture the consequences of different spatial patterns of land use and land cover change \citep{perrings2018economics}. The models developed here are a step in that direction.

\bibliography{citations,citations2}

\end{document}